\documentclass[sigconf,edbt]{acmart-edbt2019}

\usepackage{booktabs} 
\usepackage{graphicx}
\usepackage{balance}  
\usepackage{mathrsfs}
\usepackage[ruled, vlined, linesnumbered]{algorithm2e}
\usepackage{algorithmic}
\usepackage{amsmath,amssymb} 
\usepackage{subfig}
\usepackage{epsfig}
\usepackage{multirow}
\usepackage{here}
 
\newtheorem{defi}{Definition}
\newtheorem{theo}{Theorem}

\setcopyright{rightsretained}

\acmDOI{}

\acmISBN{978-3-89318-081-3}

\acmConference[EDBT 2019]{22nd International Conference on Extending Database Technology (EDBT)}{March 26-29, 2019}{Lisbon, Portugal} 
\acmYear{2019}

\begin{document}
\title{Efficient Network Reliability Computation\\ in Uncertain Graphs}
  
\author{Yuya Sasaki$^{\dagger}$, Yasuhiro Fujiwara$^{\S\dagger}$, Makoto Onizuka$^{\dagger}$}
\affiliation{%
  \institution{$\dagger$Graduate School of Information Science and Technology, Osaka University, Osaka, Japan\\
  $\S$NTT Software Innovation Center, Tokyo, Japan
  }
}
\email{sasaki@ist.osaka-u.ac.jp, fujiwara.yasuhiro@lab.ntt.co.jp, onizuka@ist.osaka-u.ac.jp}
\renewcommand{\shortauthors}{}

\begin{abstract}
Network reliability is an important metric to evaluate the connectivity among given vertices in uncertain graphs.
Since the network reliability problem is known as $\#$P-complete, existing studies have used approximation techniques.
In this paper, we propose a new sampling-based approach that efficiently and accurately approximates network reliability.
Our approach improves efficiency by reducing the number of samples based on the {\it stratified sampling}.
We theoretically guarantee that our approach improves the accuracy of approximation by using lower and upper bounds of network reliability, even though it reduces the number of samples.
To efficiently compute the bounds, we develop an extended BDD, called {\it S$^2$BDD}.
During constructing the S$^2$B\-D\-D, our approach employs dynamic programming for efficiently sampling possible graphs.
Our experiment with real datasets demonstrates that our approach is up to 51.2 times faster than existing sampling-based approach with a higher accuracy.
\end{abstract}

%
%


\maketitle

\section{Introduction}
\label{sec:intro}
To understand and design our world, we need to model and analyze relationships between objects.
Objects and relationships can be modeled by a graph, whose vertices and edges represent the objects and the relationships, respectively.
Graph analysis is widely used in many domains, and the {\it reachability} \cite{cheng2014efficient,valstar2017landmark,zhou2017dag} and {\it network reliability} \cite{ball1995network, colbourn1987combinatorics,valiant1979complexity} are the fundamental research topics in graph analysis. 
Reachability techniques compute whether there are paths between two {\it terminals} (i.e., given vertices).
On the other hand, network reliability techniques compute a probability that all pairs of terminals are connected in  {\it uncertain graphs}.
In an uncertain graph, each edge is associated with an {\it edge existence probability} to quantify the likelihood that the edge exists in the graph.
Network reliability is more generalized than reachability in terms of two aspects (1) a probabilistic value (the reachability is binary) and (2) the number of terminals.
Thus, network reliability techniques have two benefits over reachability techniques.
First, we can handle the inherent {\it uncertainty} of relationships in the real-world by modeling the uncertainty as the edge existence probability \cite{DBLP:series/ads/Aggarwal09, khan2015uncertain}.
Second, we can flexibly specify arbitrary numbers of terminals. 
From the above two benefits, the network reliability can be widely used for the uncertain graph analysis \cite{ceccarello2017clustering,zhao2014detecting} and many practical applications \cite{kalmanek2010challenges}.
For example, protein-protein interaction networks can be modeled by uncertain graphs since protein interactions are not always established due to the sensitivity to conditions \cite{asthana2004predicting,jansen2003bayesian}. 
In such protein-protein interaction networks, analysts evaluate the network reliability among several proteins as the strengths of the relationships to elucidate the functions of proteins.
The network reliability is also used in many domains such as communication networks  \cite{ball1995network,ortel1999broad} and urban planning~\cite{hamer2005value}.

Unfortunately, the computation cost of the network reliability is significantly large because it is $\#$P-complete problem \cite{valiant1979complexity}.
The high complexity of $\#$P-complete is caused by the fact that the computation of the network reliability inherently requires to enumerate all {\it possible graphs} which have the same set of vertices and an arbitrary subset of the edges without their probabilities.
Each possible graph has its probability computed from the existence probabilities of its edges.
A set of possible graphs is logically equivalent with its original uncertain graph.
To compute the network reliability, we sum up the probabilities of all  possible graphs in which all the terminals are connected.

\begin{figure}[ttt]
	\centering
	\includegraphics[width=0.9\linewidth]{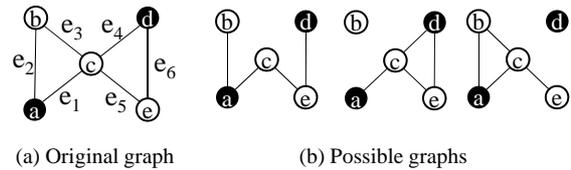}
	\vspace{-2mm}
	\caption{Uncertain graph}
	\label{fig:intro}
		\vspace{-6mm}
\end{figure}

We explain an example of computation of the network reliability by using Figure \ref{fig:intro}.
This figure shows an original uncertain graph and  three examples of its possible graphs. 
The black vertices represent terminals.
Let us assume that each edge has 0.7 as its existence probability. Since these possible graphs have four existent and two non-existent edges, their probabilities are 0.0216 (i.e., $0.7^{4} \cdot (1-0.7)^2$).
All these terminals are connected only in the left and middle possible graphs.
Thus, their probabilities are added to the network reliability.

\subsection*{Problem Definition and Technical Overview}
We approximate the network reliability since the computation cost of the network reliability is significantly large due to $\#$P-complete problem.
In this paper, we consider the problem of computing the approximate network reliability by sampling.
We formally define the problem as follows.

\vspace{1mm}
{\it Problem definition}: {\bf (Approximate network reliability)}.
Given an uncertain graph $\mathcal{G}$, a set of terminals $\mathbb{T}$, and the number of samples $s$, we efficiently compute the approximate network reliability $\hat{R}[\mathcal{G},\mathbb{T}]$.
\vspace{1mm}

The computation cost of sampling becomes considerable as the number of samples increases. 
To efficiently approximate the network reliability, we reduce the number of samples with keeping a high accuracy.
Our challenges are (1) how to reduce the number of samples with a theoretical guarantee of the accuracy and (2) how to practically achieve the theoretical results from the first challenge. 
As for the first challenge, we extend the {\it stratified sampling} \cite{thompson2002sampling}, which increases the accuracy of an estimated value by using the lower and upper bounds of the value.
We first prove a theorem that we reduce the number of samples without sacrificing the accuracy of approximation. 

We can reduce the number of samples in accordance with the theoretical results. The theoretical results have two requirements; (1) to efficiently compute the approximate network reliability, we need to efficiently obtain the tight lower and upper bounds of the network reliability and (2) to guarantee the approximation of accuracy, we need to sample possible graphs from the set of possible graphs that are not used to compute the bounds.
There are no trivial techniques to effectively achieve them.
Therefore, we develop an extended {\it binary decision diagram}, which we call {\it scalable and sampling BDD} ({\it S$^2$BDD} for short).
The S$^2$BDD enables preferentially searching for possible graphs in which terminals are connected/disconnected. The connected and disconnected possible graphs are used for computing the lower and upper bounds.
Our approach employs dynamic programming during constructing the S$^2$BDD for efficiently sampling the possible graphs. It enables avoiding sampling possible graphs from the set of possible graphs that are used to compute the bounds.

Furthermore, our approach becomes more efficient by reducing the size of graphs. Thus, we propose an extension technique of our approach which uses 2-edge connected components \cite{chang2013efficiently}.
The extension technique prunes vertices and edges that do not affect the network reliability, decomposes the graph to several subgraphs, and transforms the subgraphs into a smaller graphs.
It efficiently reduces the vertices and edges involved in the computation while preserving the network reliability.

\subsection*{Contributions and Organization}
To the best of our knowledge, our approach is the first solution to achieve both high efficiency and accuracy to compute the network reliability.
Our approach has the following attractive characteristic.
\begin{itemize}
\item Our approach improves the efficiency to compute an approximate network reliability by reducing the number of samples. The extension technique effectively reduces the size of graphs while preserving the network reliability. 
\item Our approach outputs more accurate network reliability than the existing approaches. 
We theoretically guarantee that our approach improves the accuracy of approximation, even though it reduces the number of samples.
\item Our approach computes the exact answer for small-scale graphs due to the S$^2$BDD though the existing sampling-based approach cannot compute the exact answer.
\item Our approach can be used to improve the performances on uncertain graph analyses~\cite{ceccarello2017clustering,jin2011discovering,khan2014fast} in terms of both accuracy and efficiency because many algorithms compute the network reliability by sampling techniques.
\end{itemize}
The remainder of this paper is organized as follows. Section~ \ref{sec:related} introduces related work. Section \ref{sec:preliminaries} then describes the preliminaries. Sections \ref{sec:algorithm} and  \ref{ssec:graphreduction} present our approach and an extension technique for our approach, respectively. Section \ref{sec:algo} describes algorithms of our approach with the extension.   Section \ref{sec:experiment} shows the results obtained from the experiments, and Section \ref{sec:conclusion} concludes the paper.

\section{Related work}
\label{sec:related}
Querying and mining uncertain graphs have recently attracted much attention in the database and data mining  research communities.
We review some relevant works related to the network reliability problem. 

{\bf Network reliability: }
For computing the network reliability, several approaches have been proposed such as cut-based approach and BDD-based approach.
The cut-based \cite{ahmad1988simple,harris2018improved,locks1987minimizing} approach enumerates all cuts which are divides the terminals and then computes the network reliability by using the set of cuts. 
Harris and Srinivasan \cite{harris2018improved} proposed theoretical result to obtain the lower bound of network relaibility based on cuts. However, they do not mention how to efficiently obtain the cuts.
The BDD-based approach is more efficient than the cut-based approach.
The BDD-based approach \cite{hardy2007k,maehara2017exact,yeh2002obdd} effectively avoids enumerating all possible graphs without sacrificing the exactness of the network reliability.
However, it cannot be applicable to large graphs due to the large memory usage.
The BDD-based approach first constructs a BDD, and then obtains the possible graphs in which terminals are connected by traversing the BDD.
Recent work has shown that the BDD-based approach can be applied only to graphs with 100--200 edges because of limitations of memory space \cite{hardy2007k,maehara2017exact}.
The state-of-the-art library TdZDD\footnote{https://github.com/kunisura/TdZdd} also can only be applied to very small-scale graphs.
Herrmann and Soh \cite{herrmann2009memory} proposed a memory-efficient BDD that computes the network reliability by constructing a BDD and deleting unnecessary parts of it during the process.
We partially adopt their idea to reduce the memory usage.
There are several preprocessing and indexing techniques to efficiently compute the network reliability (and similar problems) \cite{le2014improving,frey2018efficient}. 
These techniques remove redundant parts of graphs, which have similar idea of our extension technique. However, these techniques cannot directly apply to $k$-terminal reliability. 
To the best of our knowledge, there has been no prior work on approximating the network reliability with BDD.


{\bf Reachability query in uncertain graphs: } The reachability in uncertain graphs is a special type of network reliability (called $s$-$t$ network reliability)~\cite{agrawal1984time}. 
Jin et al.\ \cite{jin2011distance} proposed a distance-constraint reachability query in uncertain graphs, which answers the probability that the distance from one vertex to another is less than or equal to a threshold.
They proposed approximate algorithms as solutions to this problem.
The approximate algorithms use unequal sampling techniques \cite{rao1962simple}, and achieves higher accuracy than Monte Carlo sampling.
Cheng et al.\ \cite{cheng2016distr} proposed an algorithm to compute the reachability in distributed environments.
The algorithm reduces the size of graphs without sacrificing the exactness of the result before computing the reachability.
It divides the graph into several subgraphs and computes probabilities of the subgraphs in distributed environments.
The algorithm is only applicable to directed acyclic graphs.
While these algorithms \cite{cheng2016distr,jin2011distance} deal with uncertain graphs, their objective is to compute reachability and their algorihms cannot be applied to computing the network reliability.

{\bf Other problems with uncertain graphs: }
Many existing works in uncertain graphs use the network reliability as the metric to evaluate the connectivity among vertices. 
The efficiency and accuracy of their algorithms depend on those of the sampling techniques.
Although they use the sampling technique to compute the network reliability, they have not proposed efficient sampling techniques.
Jin et al.\ \cite{jin2011discovering} proposed an algorithm for finding reliable subgraphs in which the vertices are connected with a higher probability than a given threshold.
Ceccarello et al.\ \cite{ceccarello2017clustering} proposed clustering techniques for uncertain graphs.
The technique uses the network reliabilities between vertices as distances between them.
Khan et al.\ \cite{khan2014fast} proposed a reliability search that returns a set of vertices that are connected from given vertices with a higher probability than the threshold.
These studies have different purposes, but they use the Monte Carlo sampling to compute the network reliability.
Our approach can be used to improve their performances in terms of both accuracy and efficiency instead of using the Monte Carlo sampling.

\section{Preliminaries}
\label{sec:preliminaries}
As preliminaries of our approach, we explain uncertain graph and network reliability.
Table \ref{symbols} summarizes the notations.

\begin{table}[t]
	\begin{center}
	\caption{Notations}
	\label{symbols}
	\vspace{-3mm}
		\begin{tabular}{|l|l|} \hline
Symbol&Meaning	\\ \hline\hline
$\mathcal{G}$ & Uncertain graph \\
$\mathbb{V}$ & Set of vertices \\
$\mathbb{E}$ & Set of edges $e = (v, v')$\\
$p(e)$ & Edge existence probability of $e$ \\
$G_p$ & Possible graph \\
$\mathbb{E}_p$ & Set of edges in $G_p$\\
$Pr[G_p]$ & Existence probability of $G_p$ \\
$\mathcal{G}_{\mathbb{E}}$ & Intermediate graph \\
$\mathbb{E}_{\exists}$ & Set of existent edges in $\mathcal{G}_{\mathbb{E}}$\\
$\mathbb{E}_{\lnot}$ & Set of non-existent edges in $\mathcal{G}_{\mathbb{E}}$\\
$Pr[\mathcal{G}_{\mathbb{E}}]$ & Existence probability of $\mathcal{G}_{\mathbb{E}}$ \\
$\mathbb{T}$ & Set of terminals\\
$R[\mathcal{G},\mathbb{T}]$  & Network reliability of $\mathcal{G}$ for $\mathbb{T}$ \\
$\hat{R}[\mathcal{G},\mathbb{T}]$  & Approximate network reliability of $\mathcal{G}$ for $\mathbb{T}$ \\
$k$ & The number of terminals \\
$w$ & Maximum size of BDD\\
$\mathbb{F}_l$ & Set of frontiers at layer $l$\\
$|\cdot|$ & The number of elements in  a set \\
\hline
		\end{tabular}
				\vspace{-4mm}
        \end{center}
\end{table}

\subsection{Uncertain graph}
Let  $\mathcal{G} =(\mathbb{V}, \mathbb{E}, p)$ be a connected and undirected uncertain graph, where $\mathbb{V}$ is a set of vertices, $\mathbb{E} \subseteq \mathbb{V} \times \mathbb{V}$ is a set of {\it uncertain} edges, and $p :\mathbb{E} \rightarrow (0,1]$ is a function that determines the edge existence probability $p(e)$ of uncertain edge $e \in \mathbb{E}$ in the graph.
We denote edge $e \in \mathbb{E}$ between $v$ and $v'$ as $e =(v,v')$.
A state of uncertain edge $e$ is {\it existent} with a probability $p(e)$ or {\it non-existent} with a probability $(1-p(e))$.
We assume that edge existence probabilities of different edges are independent of one another \cite{ceccarello2017clustering,jin2011distance}.

A {\it possible graph} $G_p =(\mathbb{V}, \mathbb{E}_p)$ is a graph that contains a set of vertices and a subset of edges of  $\mathcal{G}$ without their edge existence probabilities.
Edges in $\mathbb{E} \backslash \mathbb{E}_p$ are non-existent in the possible graph.
Although edges in possible graphs have no probabilities, the possible graphs themselves have existent probabilities.
The existent probability $Pr[G_p]$ of possible graph $G_p$ is as follows:
\begin{equation}\textstyle
Pr[G_p] = \prod_{e \in \mathbb{E}_p} p(e) \cdot \prod_{e \in \mathbb{E} \backslash \mathbb{E}_p} (1 - p(e)).\nonumber
\end{equation}
The total number of the possible graphs of $\mathcal{G}$ is $2^{|\mathbb{E}|}$ because each edge is either existent or non-existent.
We define $\mathbb{W}^{\mathcal{G}}$ as all possible graphs obtained from $\mathcal{G}$.

We define an {\it intermediate graph} $\mathcal{G}_{\mathbb{E}}(\mathbb{E}_{\exists}, \mathbb{E}_{\lnot})$, which is an uncertain graph with the set of existent edges $\mathbb{E}_{\exists}$, the set of non-existent edges $\mathbb{E}_{\lnot}$, and the set of uncertain edges $\mathbb{E} \backslash (\mathbb{E}_{\exists} \cup  \mathbb{E}_{\lnot})$.
The existent probability $Pr[\mathcal{G}_{\mathbb{E}}(\mathbb{E}_{\exists}, \mathbb{E}_{\lnot})]$ of the intermediate graph $\mathcal{G}_{\mathbb{E}}(\mathbb{E}_{\exists}, \mathbb{E}_{\lnot})$ is as follows:
\begin{equation}\textstyle
Pr[\mathcal{G}_{\mathbb{E}}(\mathbb{E}_{\exists}, \mathbb{E}_{\lnot})] = \prod_{e \in \mathbb{E}_{\exists}} p(e) \cdot \prod_{e \in \mathbb{E}_{\lnot}} (1 - p(e)).\nonumber
\end{equation}
We simply use $Pr[\mathcal{G}_{\mathbb{E}}]$ as $Pr[\mathcal{G}_{\mathbb{E}}(\mathbb{E}_{\exists}, \mathbb{E}_{\lnot})]$.
We define $\mathbb{W}^{\mathcal{G}_{\mathbb{E}}}$ as all possible graphs obtained from  $\mathcal{G}_{\mathbb{E}}$.
The total number of the possible graphs of $\mathcal{G}_{\mathbb{E}}(\mathbb{E}_{\exists}, \mathbb{E}_{\lnot})$ is $2^{|\mathbb{E} \backslash (\mathbb{E}_{\exists} \cup  \mathbb{E}_{\lnot})|}$.
We define that vertices are connected in intermediate graphs if there are paths among the vertices by existent edges, and  vertices are disconnected if there are no paths among the vertices by existent and uncertain edges.
Note that it is unsure to be connected or disconnected even if there are paths among the vertices by uncertain edges.

\subsection{Network reliability}
\label{ssec:newtwork}

The network reliability is computed by summing up the probabilities of all possible graphs in which all terminals (a subset of vertices) are connected. The definition is as follows:

\begin{defi}[Network reliability]\label{def:network} Given a set of $k$ terminals $\mathbb{T}$ and an uncertain graph $\mathcal{G}$, the network reliability $R[\mathcal{G},\mathbb{T}]$ is 
\begin{equation}\textstyle
R[\mathcal{G},\mathbb{T}] = \sum_{G_p \in  \mathbb{W}^{\mathcal{G}}} I(G_p, \mathbb{T}) \cdot Pr[G_p],
\end{equation}
where $G_p$ denotes a possible graph, and $I(G_p, \mathbb{T})$ is an indicator function that returns one if all terminals in $\mathbb{T}$ are connected in $G_p$, and returns zero, otherwise.
\end{defi}
We denote by $\hat{R}[\mathcal{G},\mathbb{T}]$ the approximate network reliability.
We simply use $R$ and $\hat{R}$ as $R[\mathcal{G},\mathbb{T}]$ and $\hat{R}[\mathcal{G},\mathbb{T}]$ for the given uncertain graph and terminals, respectively.

The network reliability with $k$ terminals is called the {\it $k$-terminal reliability}, and it is known as the most generalized network reliability \cite{hardy2007k}.
The network reliability problem is $\#$P-complete  \cite{valiant1979complexity}. 
Planar graphs can be more efficiently solved than general graphs, but it is also $\#$P-complete \cite{provan1986complexity}.
Therefore, it has no polynomial time algorithm unless $P=NP$.

BDD \cite{hardy2007k} and sampling \cite{jin2011distance} are main techniques to compute the network reliability.
BDD-based approach can compute the exact answer in small-scale graphs, while sampling-based appraoch can compute approximate answers in large-scale graphs.

\subsubsection{Binary decision diagram}\label{sssec:bdd}
\begin{figure}[ttt]
	\centering
	\includegraphics[width=0.8\linewidth]{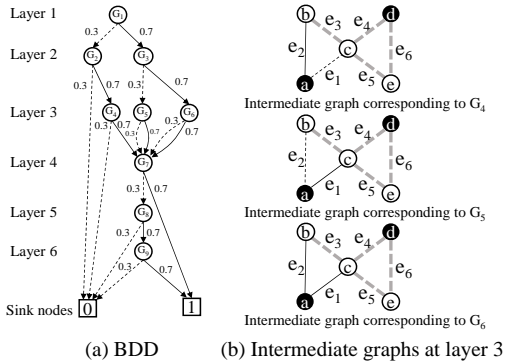}
	\vspace{-2mm}
	\caption{BDD for the original graph on Figure \ref{fig:intro}(a).}
	\label{fig:bdd}
	\vspace{-3mm}
\end{figure}

A BDD  $\mathcal{D} =(\mathbb{N}, \mathbb{A})$ is a directed acyclic graph with sets of nodes $\mathbb{N}$ and arcs $\mathbb{A}$\footnote{To avoid confusion, we use the terms ``vertex'' and ``edge'' to refer to a vertex and an edge in an uncertain graph, respectively, and ``node'' and ``arc'' to refer to a vertex and an edge in a BDD, respectively.}.
Figure \ref{fig:bdd}(a) shows the BDD to compute the network reliability of the original graph in Figure~\ref{fig:intro}.
Nodes in the BDD correspond to intermediate graphs, and arcs in the BDD correspond to existent/non-existent edges.
The BDD has a single node that has no incoming arcs, called the {\it root node} (node $G_1$ in Figure \ref{fig:bdd}(a)).
Each node has two outgoing arcs, called the {\it 0-arc} and {\it 1-arc} (represented by dashed and solid arrows in Figure \ref{fig:bdd}(a), respectively).
0-arcs and 1-arcs indicate that edges are non-existent and existent in the uncertain graph, respectively.
Each arc is associated with a {\it weight} that represents the existent or non-existent probability of the edge. 
We define {\it layer} $l$ $(\geq 1)$ as the depth from the root node.
The nodes at layer $l$ of the BDD correspond to the intermediate graphs whose edges $e_1, \ldots, e_{l-1}$ are existent/non-existent and the other edges $e_{l},\ldots, e_{|\mathbb{E}|}$ are uncertain. 
The BDD has special nodes that have no outgoing arcs, called {\it sink nodes}.
The sink nodes are of two types, called {\it 1-sink} and {\it 0-sink} (represented by rectangles with labels 1 and 0 in Figure \ref{fig:bdd}(a), respectively). 
If the terminals in the intermediate graph are connected and disconnected, the arcs point at the 1-sink and 0-sink, respectively.
We can obtain intermediate graphs in which terminals are connected by traversing the BDD from the root node to the 1-sink.

To construct the BDD, the {\it frontier-based method} is  a common procedure \cite{kawahara2017frontier,maehara2017exact}.
This method first orders edges ($e_1,\ldots, e_{|\mathbb{E}|}$).
It generates the nodes on layer $l+1$ by setting the states of $e_l$ when a BDD is already constructed until layer $l$.
In the frontier-based method, a vertex that has both existent/non-existent and uncertain edges are called a {\it frontier} $f$, and we denote by ${\mathbb{F}}_l$ the set of frontiers at layer $l$.
Figure~\ref{fig:bdd}(b) shows intermediate graphs after processing $e_1$ and $e_2$, where solid black, dashed black, and dashed gray lines denote existent, non-existent, and uncertain edges, respectively. These intermediate graphs correspond to $G_4$, $G_5$, and $G_6$ in the BDD from the top, respectively. 
Vertices $b$ and $c$ are frontiers because they have both existent/non-existent and uncertain edges.
Note that nodes at the same layer $l$ have the same set of frontiers $\mathbb{F}_l$.
The frontier-based method maintains several attributes on only the frontiers (e.g., the number of uncertain edges and the number of terminals connected to the frontiers).
It merges the nodes if the attributes are the same.
Thus, the frontier-based method can effectively reduce the number of nodes.

The size of the BDD is defined by the number of nodes in the BDD \cite{hardy2007k}. 
Generally, it exponentially increases as the number of edges in the uncertain graphs increases.
As the size of the BDD increases, both of the computation cost and the memory usage increase.
Thus, it is hard to compute the exact network reliability. 

\subsubsection{Sampling}\label{sssec:sampling}
Sampling is a basic approach for computing the approximate network reliability \cite{cheng2016distr,jin2011discovering,jin2011distance}.
Given the number of samples~$s$, the sampling-based approach repeats the following procedures $s$ times: (1) picking a possible graph of $\mathcal{G}$ as a sample, $G_{p_i}~(1\leq i \leq s)$ according to the probabilities $Pr[G_{p_i}]$ from $\mathbb{W}^{\mathcal{G}}$ and then (2) computing whether all the terminals are connected or not in $G_{p_i}$.
The time complexity of the sampling-based approach is $O(s \cdot ( |\mathbb{V}|  + |\mathbb{E}|))$.
This is because it requires $O(|\mathbb{E}|)$ time to determine the states of all edges and $O(|\mathbb{V}|+|\mathbb{E}|)$ time to compute the connectivity by a depth first search for each sample.

The accuracy of the sampling-based approach is evaluated by its variance. 
Since the sampling-based approach is a randomized algorithm \cite{motwani2010randomized}, the average network reliability is most likely to be closest to the exact network reliability.
A small variance indicates a small rate of error (i.e., high accuracy).
Note that {\it unbiased sampling} is necessary that samples possible graphs according to their probabilities for guaranteeing the theoretical variance.
As the number of samples increases, the variance decreases but the computation cost increases.
Therefore, there is a trade-off between the accuracy and the computation cost.


The {\it stratified sampling} is known as a successful method in the field of statistics \cite{thompson2002sampling}.
The stratified sampling divides the population into subgroups and individually picks samples from each subgroup.
The variance of the estimated value for the whole population are the sum of the variances of the estimated values for individual subgroups.
Let $L$ be the number of subgroups and $R_i$ be the estimated total probabilities of possible graphs for subgroup $i$.
The estimated network reliability is computed by summing up the total probabilities for the subgroups as follows:
\begin{equation}\textstyle
\hat{R} = \sum_{i=1}^L \hat{R_i}. \nonumber
\end{equation}
The variance is the sum of the individual variances for the subgroups as follows:
\begin{equation}\textstyle
Var[\hat{R}] = \sum_{i=1}^L Var[\hat{R_i}]. \nonumber
\end{equation}
When we compute the exact values for the subgroups, the variances of the estimated network reliability for the subgroup become zero.
Thus, when we compute the exact values for the subgroups, the variance of the estimated network reliability for the whole population decreases.

\section{Our Approach}
\label{sec:algorithm}
In this paper, we solve the problem of the approximate network reliability.
Section \ref{ssec:highleveloverview} provides an overview of our approach.
Section \ref{ssec:sampling} explains how to reduce the number of samples.
Section \ref{ssec:r2bdd} presents our extended BDD S$^2$BDD.

\subsection{Overview}
\label{ssec:highleveloverview}

Our approach efficiently and accurately computes the approximate network reliability.
We achieve high efficiency and accuracy with the following ideas:
\begin{itemize}
\item {\bf Reduction of the number of samples}: Our approach significantly reduces the number of samples with keeping a high accuracy of approximation by using the lower and upper bounds of the network reliability. 
\item {\bf Efficient computation of the bounds of network reliability}: We develop the S$^2$BDD to efficiently compute the bounds of the network reliability.  
\item {\bf Dynamic programming}:  During constructing S$^2$BDD, we employ dynamic programming for efficiently sampling possible graphs.
\end{itemize}

Our approach reduces the number of samples in accordance with the stratified sampling. 
We theoretically guarantee that the number of samples becomes small as the lower and upper bounds become tight without sacrificing the accuracy of approximation.
We prove it in two representative estimators; Monte Carlo and Horvitz-Thompson estimators \cite{thompson2002sampling}.

For achieving the theoretical result, we compute the lower and upper bounds by constructing the S$^2$BDD. 
We specify the maximum size $w$ of S$^2$BDD for avoiding a large cost to construct the S$^2$BDD.
Our approach deletes nodes on the S$^2$BDD when its size exceeds $w$.
To effectively delete nodes, we define a heuristic function for preferentially keeping {\it high-priority} nodes in the S$^2$BDD; the priorities are computed from the possibilities of improving the bounds.
The S$^2$BDD enables efficiently computing the bounds because nodes preferentially point at sink nodes.


For efficiently sampling possible graphs, our approach employs dynamic programming during constructing the S$^2$BDD. 
We can straightforwardly employ dynamic programming for sampling because sampling possible graphs from intermediate graphs is a sub problem of sampling possible graphs from the original uncertain graph.
We also use the {\it stratified random sampling} for determining the number of samples for each sub problem. 
The stratified random sampling divides the set of possible graphs into subgroups and samples possible graphs from each subgroup.
\subsection{Reducing the number of samples}
\label{ssec:sampling}

In this section, we theoretically prove that our approach reduces the number of samples while keeping a high accuracy in accordance with the stratified sampling \cite{fishman1986comparison,manzi2001fishman}.
As we mentioned in Section \ref{sssec:sampling}, the accuracy of sampling is evaluated by the variance of the estimated network reliability.
Since the stratified sampling reduces the variance of the estimated network reliability, we can reduce the number of samples without sacrificing the accuracy of approximation.


To apply the stratified sampling, we divide the set $\mathbb{W}^{\mathcal{G}}$ of possible graphs into three subgroups $\mathbb{W}_c^{\mathcal{G}}$, $\mathbb{W}_d^{\mathcal{G}}$, and $\mathbb{W}_u^{\mathcal{G}}$.
$\mathbb{W}_c^{\mathcal{G}}$ and $\mathbb{W}_d^{\mathcal{G}}$ include the sets of only possible graphs in which terminals are connected and disconnected, respectively.
$\mathbb{W}_u^{\mathcal{G}}$ includes the set of possible graphs that are not included in $\mathbb{W}_c^{\mathcal{G}}$ and $\mathbb{W}_d^{\mathcal{G}}$.
Let $p_c$ and $p_d$ be the sum of the probabilities of possible graphs in $\mathbb{W}_c^{\mathcal{G}}$ and $\mathbb{W}_d^{\mathcal{G}}$, respectively. 
Hence, from Definition \ref{def:network}, the upper and lower bounds are given as follows:
\begin{eqnarray}\textstyle
\!\!\!R \!\!\!&=&\textstyle \!\!\!\! \sum_{G_p \in  \mathbb{W}_c^{\mathcal{G}}} Pr[G_p] + \sum_{G_p \in  \mathbb{W}_u^{\mathcal{G}}} I(G_p, \mathbb{T}) Pr[G_p] \nonumber\\
&=&\textstyle  \!\!\!\! p_c + \sum_{G_p \in  \mathbb{W}_u^{\mathcal{G}}} I(G_p, \mathbb{T}) Pr[G_p] \nonumber\\
&\geq& \!\!\!\! p_c. \nonumber\\
\!\!\!R \!\!\!&=&\textstyle \!\!\!\! 1 - \sum_{G_p \in  \mathbb{W}_d^{\mathcal{G}}} Pr[G_p] - \sum_{G_p \in  \mathbb{W}_u^{\mathcal{G}}}( Pr[G_p] - I(G_p, \mathbb{T}) Pr[G_p])  \nonumber\\
 &=&\textstyle \!\!\!\! 1 - p_d - \sum_{G_p \in  \mathbb{W}_u^{\mathcal{G}}}( Pr[G_p] - I(G_p, \mathbb{T}) Pr[G_p]) \nonumber\\
&\leq& \!\!\!\! 1-p_d. \nonumber
\end{eqnarray}
Consequently, we have  $p_c \leq R \leq 1-p_d$.
We reduce the number of sample by using the lower bound $p_c$ and upper bound $1-p_d$.

The variance also depends on estimators.
In our approach, we exploit two representative estimators; Monte Carlo estimator and Horvitz-Thompson estimator.
The Monte Carlo estimator is a basic technique for computing the average values of the samples. 
On the other hand, the Horvitz-Thompson estimator is unequal probability estimator, which provides smaller variance than the Monte Carlo estimator under sampling without replacement.
We explain how to reduce the number of samples in the two estimators with keeping  a high accuracy. 

\noindent
{\bf Monte Carlo estimator: }
The Monte Carlo estimator for $R$ is:
\begin{equation}\textstyle
\hat{R} = \frac{\sum_{i=1}^{s} I(G_{p_i},\mathbb{T})}{s}. \nonumber
\end{equation}
The variance is computed by the following equation \cite{fishman1986comparison}:
\begin{equation}\textstyle
Var[\hat{R}] = \frac{R(1-R)}{s}. \nonumber
\end{equation}

Because the random sampling is unbiased, i.e., $E(\hat{R}) = R$, the variance can be simply written as follows \cite{manzi2001fishman}:
\begin{equation}\label{eq:monte1}\textstyle
Var[\hat{R}] = \frac{R(1-R)}{s} \approx  \frac{\hat{R}(1-\hat{R})}{s}.
\end{equation}

Let $Var[\hat{R}]' $ be the variance using the upper and lower bounds. $Var[\hat{R}]' $ is computed  in accordance with the stratified sampling as follows \cite{fishman1986comparison,manzi2001fishman}:
\begin{equation}\label{eq:monte2}\textstyle
Var[\hat{R}]' =  \frac{(\hat{R}-p_c)(1 - p_d - \hat{R})}{s}.
\end{equation}
From Equations (\ref{eq:monte1}) and (\ref{eq:monte2}),  we obtain the following equation:
\begin{equation}\label{eq:monte_v}\textstyle
\frac{\hat{R}(1-\hat{R})}{s} \geq \frac{(\hat{R} - p_c)(1 - p_d - \hat{R})}{s}.
\end{equation}
Therefore, we have $Var[\hat{R}] \geq Var[\hat{R}]'$.
From Equation (\ref{eq:monte_v}), we obtain the following theorem:

\begin{theo}\label{theo:sample1}
Given the number of samples $s$, the lower bound $p_c$, and the upper bound $1 - p_d$, the variance of network reliability by using Monte Carlo estimator with $s'~(\leq s)$ samples is less than and equal to that with $s$ samples if $s'$ is computed by the following equations: 
\[
  s' = \begin{cases}
    \lfloor s  (1-p_d) \rfloor. & (p_c=0) \\
\lfloor s  (1-p_c) \rfloor. & (p_d=0) \\
\lfloor s  (1-4\cdot p_c (1-p_c)) \rfloor. & (p_c=p_d) \\
\lfloor s  (1-4\cdot p_c (1-p_d)) \rfloor. & (p_c<p_d) \\
\lfloor s  (1- \mathrm{min} ( 4 p_c (1-p_c),	& \\
 ~~~~~ 4 (p_c(1-p_d) + (p_d-p_c)) ) \rfloor. & (p_c>p_d) 
  \end{cases}
\]
\end{theo}
{\it Proof: } From Equation (\ref{eq:monte_v}), we have the following equation such that the variance with $s$ samples is equal to that with $s'$ samples by using the lower and upper bounds:
\begin{equation} \textstyle
\frac{(p_c-\hat{R})(1 - p_d - \hat{R})}{s'} =  \frac{\hat{R}(1-\hat{R})}{s} \nonumber
\end{equation}
Then, $s'$ is computed as follows:
\begin{eqnarray}\label{eq:sample_theo}\textstyle
s' &=& \textstyle s \cdot \frac{(\hat{R}-p_c)(1 - p_d - \hat{R})}{\hat{R}(1-\hat{R})} \nonumber \\ 
	&=& \textstyle s \cdot \left(1 - \frac{p_c(1-\hat{R})+ p_d(\hat{R}-p_c)}{\hat{R}(1-\hat{R})}\right)
\end{eqnarray}
However, we cannot compute $\hat{R}$ before sampling $s$ possible graphs.
Therefore, we remove $\hat{R}$ from Equation (\ref{eq:sample_theo}) by dividing the patterns of $p_c$ and $p_d$.
First, if $p_c=0$, $s'$ is computed as follows:
\begin{eqnarray}\textstyle
& s  \left(1 - \frac{p_d  \hat{R}}{\hat{R}(1-\hat{R})}\right) \leq s  (1-p_d).  \nonumber\\ \nonumber
& s' =\lfloor  s  (1-p_d) \rfloor.
\end{eqnarray}
Second, if $p_d=0$, $s'$ is computed as follows:
\begin{eqnarray}\textstyle
& s  \left(1 - \frac{p_c  (1-\hat{R})}{\hat{R}(1-\hat{R})}\right) \leq s (1-p_c). \nonumber\\ \nonumber
& s' =\lfloor  s  (1-p_c) \rfloor.
\end{eqnarray}
Third, if $p_c=p_d$, $s'$ is computed as follows: 
\begin{eqnarray}\textstyle
&\label{eq:pcpd} s  \left(1 - \frac{p_c  (1-\hat{R}) + p_c  (\hat{R}-p_c)}{\hat{R}(1-\hat{R})}\right) \leq s (1-4p_c(1-p_c)). \\ \nonumber
& s' =\lfloor  s (1-4p_c(1-p_c)) \rfloor.
\end{eqnarray}
In Equation (\ref{eq:pcpd}), the maximum value of $\hat{R}(1-\hat{R})$ is 0.25. Thus, we substitute 0.25 for $\hat{R}(1-\hat{R})$ in the denominator.
Fourth, if $p_c<p_d$,  $s'$ is computed as follows: 
\begin{eqnarray}\textstyle
& s  \left(1 - \frac{p_c  (1-\hat{R}) + p_d  (\hat{R}-p_c)}{\hat{R}(1-\hat{R})}\right) \leq s (1-4p_c(1-p_d)). \nonumber\\ \nonumber
& s' =\lfloor  s (1-4p_c(1-p_d)) \rfloor.
\end{eqnarray}
Finally, if $p_c>p_d$, $s'$ is computed as follows: 
\begin{eqnarray}\textstyle
& \textstyle s  \left(1 - \frac{p_c  (1-\hat{R}) + p_d  (\hat{R}-p_c)}{\hat{R}(1-\hat{R})}\right) \leq s (1-4p_c(1-p_c)).\nonumber\\
&\!\!\!\!\!\!\! \textstyle s  \left(1 - \frac{p_c  (1-\hat{R}) + p_d  (\hat{R}-p_c)}{\hat{R}(1-\hat{R})}\right) \leq s (1-4(p_c(1-p_c)+(p_d-p_c)).\nonumber\\
\!\!\!\!\!\!\!\!\!\!\!\!\!\!&\!\!\!\!\!\!\!\!\!\!\!\!\!\!\label{eq:pcbpd} \textstyle s' =\lfloor   s  (1- \mathrm{min} (4 p_c (1-p_c), 4 (p_c(1-p_d) + (p_d-p_c)) )) \rfloor.
\end{eqnarray}
In Equation (\ref{eq:pcbpd}), the minimum $s'$ depends on the values of $p_c$ and $p_d$.
Consequently, we have that $s' \leq s$ for all patterns of $p_c$ and $p_d$. 
\hfill{} $\square$
\vspace{1mm}

\noindent
{\bf Horvitz-Thompson estimator: }
The Horvitz-Thompson estimator  for $R$ is:
\begin{equation}\textstyle
\hat{R} = \frac{\sum_{i=1}^{s} Pr[G_{p_i}] \cdot I(G_{p_i},\mathbb{T})}{\pi_i},\nonumber
\end{equation}
where $\pi_i = 1-(1-Pr[G_{p_i}])^s$.
The variance is:
\begin{eqnarray}
 \textstyle Var[\hat{R}] &\!\!\!\!\!=\!\!\!\!\!\!&  \textstyle \sum_{i=1}^s \left( \frac{1-\pi_i}{\pi_i} \right)  I(G_{p_i},\mathbb{T}) Pr[G_{p_i}]^2\nonumber \\
& \textstyle\!\!\!\!\!\!\!\!\!\!\!\!& \textstyle\!\!\!\!\!\!\!\!\!\!\!\!\!\!\!\!  +\sum_{i}^s \sum_{j,i\neq j}^s \!\!\left( \frac{\pi_{ij} -\pi_i \pi_j}{\pi_i \pi_j} \right) \! I(G_{p_i},\!\mathbb{T})  I(G_{p_j},\!\mathbb{T})Pr[G_{p_i}] Pr[G_{p_j}], \nonumber
\end{eqnarray}
where  $\pi_{ij} = 1-(1-Pr[G_{p_i}])^s - (1-Pr[G_{p_j}])^2 + (1-Pr[G_{p_i}]-Pr[G_{p_j}])^s$.
The variance is simplified as follows~\cite{jin2011distance}:
\begin{equation}\label{eq:sample2.1}\textstyle
Var[\hat{R}] = \frac{R(1-R)}{s} -  \frac{\Sigma_{i=1}^s (s-1)   I(G_{p_i},\!\mathbb{T}) Pr[G_{p_i}]^2}{2s}.
\end{equation}

The variance using the lower and upper bounds is computed  in accordance with the stratified sampling as follows:
\begin{equation}\label{eq:sample2.2}\textstyle
Var[\hat{R}]' =  \frac{(\hat{R}-p_c)(1 - p_d - \hat{R})}{s}  - \frac{\sum_{i=1}^s (s-1)  I(G_{p_i},\mathbb{T}) Pr[G_{p_i}]^2}{2s}.
\end{equation}

\begin{theo}\label{theo:sample2}
Given the number of samples $s$, the lower bound $p_c$, and the upper bound $1 - p_d$, the variance of network reliability by using Horvits-Thompson estimator with $s'~(\leq s)$ samples is less than and equal to that with $s$ samples where $s'$ is equal to the number of samples in Monte Carlo estimator in \ref{theo:sample1}.
\end{theo}
{\it Proof: }  From Equations (\ref{eq:sample2.1}) and (\ref{eq:sample2.2}), we have the following equation: 
\begin{eqnarray} \textstyle
& &\!\!\!\!\!\!\!\!\!\!\! \textstyle\frac{(\hat{R}-p_c)(1 - p_d - \hat{R})}{s'}  -  \frac{\sum_{i=1}^s (s'-1)   I(G_{p_i},\mathbb{T}) Pr[G_{p_i}]^2}{2s'}  \nonumber\\
& & \textstyle~~~~~~~=  \frac{\hat{R}(1 - \hat{R})}{s}  -  \frac{\sum_{i=1}^s (s-1)   I(G_{p_i},\mathbb{T}) Pr[G_{p_i}]^2}{2s}. \nonumber
\end{eqnarray}
The values of the right are the same because the estimator is unbiased.
The proof for this follows Theorem \ref{theo:sample1}. \hfill{} $\square$
\vspace{1mm}

Our approach reduces the number of samples in accordance with Theorems \ref{theo:sample1} and \ref{theo:sample2}.
As a result, our approach is more efficient than the existing sampling-based approach.

\subsection{Scalable and Sampling BDD: S$^2$BDD}
\label{ssec:r2bdd}
We can reduce the number of samples by using the lower and upper bounds of network reliability.
To efficiently obtain the bounds, we develop the S$^2$BDD.
We efficiently search for the possible graphs in which terminal are connected and disconnected with high probabilities by constructing the S$^2$BDD.
Furthermore, during constructing the S$^2$BDD, we sample possible graphs that are not used to compute the bounds, which is the requirement of stratified sampling.
Our approach uses S$^2$BDD for both computing the bounds of network reliability and sampling possible graphs. 

We design the S$^2$BDD to effectively reduce its size.
The S$^2$BDD keeps a single layer and sink nodes while ordinary BDD contains all layers.  
This idea is based on the observation that the layer $l-1$ is unnecessary after constructing the next layer $l$ to both construct the layer $l+1$ and obtain the bounds.
We first define the S$^2$BDD and then explain how to construct it.
\begin{defi}\label{def:s2bdd} Let $\mathbb{N}_l$ be a set of nodes at layer $l$.  S$^2$BDD consists $\mathbb{N}_l$, the 1-sink, and the 0-sink.
The S$^2$BDD maintains the following attributes on node $n \in \mathbb{N}$:
\begin{itemize}
\item $p_n$: the probability of the intermediate graph corresponding to node $n$.
\item $\{c_{n,f}\}$ for all $f \in \mathbb{F}_l$: an identifier of connected component. If frontiers $f$ and $f'$ $\in \mathbb{F}_l$ are connected by existent edges, $c_{n,f}$ and $c_{n,f'}$ share the same identifier.
\item $\{d_{n,f}\}$ for all $f \in \mathbb{F}_l$: the sum of the numbers of uncertain edges connected to the frontiers such that $\{ f' \in \mathbb{F}_l | c_{n,f}=c_{n,f'} \}$. 
\item $\{t_{n,f}\}$ for all $f \in \mathbb{F}_l$: the number of the terminals that are connected  to $f$ by existent edges. 
\end{itemize}
The 1-sink and 0-sink maintain the probabilities $p_c$ and $p_d$ that terminals are connected and disconnected, respectively. 
\end{defi}
For example, in Figure \ref{fig:bdd}, S$^2$BDD contains third and sink layers but does not contains first and second layers. 


To construct an S$^2$BDD, we process edge $e_l$ and generate the set of nodes $\mathbb{N}_{next}$ at layer $l+1$.
The construction method comprises four procedures; {\it generating}, {\it merging}, {\it deleting}, and {\it sampling}.
The following sections explain these procedures in details.

\subsubsection{Generating and Merging Procedures}
\label{sssec:generate}
The BDD-based approach uses the generating and merging procedures to construct the BDD.
We extend these procedures to effectively compute the bounds without sacrificing the exactness of the network reliability.
For extending the generating and merging procedures, we capture the feature of computing the network reliability such that we can skip the computation of nodes when we obtain the probabilities $p_c$ and $p_d$ exactly.

We first explain the generating procedure.
The generating procedure sets the state of edge $e_l$ (recall that arcs at layer $l$ in the BDD corresponding to $e_l$) and then generates the set of new nodes $\mathbb{N}_{next}$ at layer $l+1$.
As the same as the traditional procedure, we generate two new nodes at layer $l+1$ from every node at layer $l$ according to the state of $e_l$.
We set the attributes on the new nodes (i.e., $p_n$, $\{c_{n,f}\}$, $\{d_{n,f}\}$, and $\{t_{n,f}\}$).
More specifically, $p_n$ is set as $p_n \cdot p(e_l)$ when $e_l$ is existent and set as $p_n \cdot (1-p(e_l))$ when $e_l$ is non-existent.
$\{c_{n,f}\}$, $\{d_{n,f}\}$, and $\{t_{n,f}\}$ are computed from attributes of frontiers on nodes at layer $l$ by merging attributes of frontiers and creating new frontiers.
If all the terminals in the intermediate graph are connected, we add its probability to $p_c$, and if they are disconnected, we add its probability to $p_d$.

If we determine whether or not terminals are c\-o\-n\-n\-e\-c\-t\-e\-d/d\-i\-s\-c\-o\-n\-n\-e\-c\-t\-e\-d with processing a smaller number of edges, we can obtain the tight bounds of the network reliability earlier.
Let $n$, $n'$, $\mathbb{F}$, and $\mathbb{F}'$ be the new node at layer $l+1$, the node before setting $e_l$ of $n$ at layer $l$, the sets of frontier at layers $l+1$ and $l$, respectively.
We determine whether or not terminals are c\-o\-n\-n\-e\-c\-t\-e\-d/d\-i\-s\-c\-o\-n\-n\-e\-c\-t\-e\-d based on following lemmas:

\begin{lemma}\label{lemma:connect}
All the terminals $t \in \mathbb{T}$ are connected if the attributes of the frontiers satisfy one of the following conditions:\\
{\bf Condition 1}: edge $e_l=(v,v')$ is existent, for $t_{n, f} = k$, $\exists f \in \mathbb{F}$.\\
{\bf Condition 2}: edge $e_l=(v,v')$ is existent, for (1) $v \in \mathbb{F}'$, (2) $v' \not\in \mathbb{F}' \cup \mathbb{F}$, (3) $t_{n',v} = k-1$, and (4)  $v \in \mathbb{T}$ (similarly,  replacing $v$ with $v'$ and vice versa).\\
{\bf Condition 3}: edge $e_l=(v,v')$ is existent, for (1) $v, v' \in \mathbb{F}$, (2) $c_{n',v} \neq c_{n',v'}$, and (3) $t_{n',v}+t_{n',v'} = k$.
\end{lemma}
\vspace{-1mm}
{\it Proof:} This is an immediate consequence of the definitions because all the terminals are connected.
 \hfill{} $\square$ 
 
\begin{lemma}\label{lemma:disconnect}
The terminals are disconnected if the attributes of the frontiers satisfy one of the following conditions:\\
{\bf Condition 1}: edge $e_l=(v,v')$ is non-existent, for (1) $v \not\in \mathbb{F}' \cup \mathbb{F}$, and (2) $v \in \mathbb{T}$ (similarly,  for $v'$).\\
{\bf Condition 2}: edge $e_l=(v,v')$ is non-existent, for (1) $v \in \mathbb{F}'$, (2) $t_{n',v}>0$, and (3) $d_{n',v}\!=\!1$ (similarly, for $v'$).\\
{\bf Condition 3}: edge $e_l=(v,v')$ is existent or non-existent, for (1) $v, v' \in \mathbb{F}'\backslash \mathbb{F}$ and (2) ($t_{n',v}>0$ or $t_{n',v'} >0$).
\end{lemma}
\vspace{-1mm}
{\it Proof:} This is an immediate consequence of the definitions because the terminals are disconnected.
 \hfill{} $\square$ 
\vspace{1mm}

\noindent
Note that the state-of-the-art construction of the BDD uses only the condition 1 on Lemmas 1 and 2.
As a result, the S$^2$BDD can more effectively tighten the bounds of network reliability.

We next explain the merging procedure.
Since each intermediate graph on S$^2$BDD has different existent and non-existent edges, the attributes on each frontier are different (in general).
The merging procedure merges the nodes that make a transition to the same sink nodes based on the following lemma:

\begin{lemma}\label{lemma:merge}
Given nodes $n_1$ and $n_2$ at layer $l$, if we have  for $\forall  f \in \mathbb{F}_l$ (1) $c_{n_1,f} = c_{n_2,f}$ and (2) ($t_{n_1,f}= 0$ and $t_{n_2,f}=0$) or ($t_{n_1,f}> 0$ and $t_{n_2,f}>0$), then nodes derived from $n_1$ and $n_2$ with the same states of edges $e_{l+1},\ldots, e_{|\mathbb{E}|}$ make a transition to the same sink nodes.
\end{lemma}
\vspace{-1mm}
{\it Proof:} If $n_1$ and $n_2$ have (1) $\{c_{n_1,f}\}$ = $\{c_{n_2,f}\}$ for all $f$ in $\mathbb{F}_l$, the connected frontiers are the same in the intermediate graphs corresponding to $n_1$ and $n_2$.
New nodes $n_1'$ and $n_2'$ derived from $n_1$ and $n_2$ are the same $\{c_{n_1',f}\} = \{c_{n_2',f}\}$ if they have the same states of edges $e_{l+1},\ldots, e_{|\mathbb{E}|}$.
Thus,  $\{c_{n_1,f}\}$ and $\{c_{n_2,f}\}$ for all $f$ in $\mathbb{F}_l$ are the same until they make a transition to the sink nodes.
Since the same $\{c_{n_1,f}\}$ and $\{c_{n_2,f}\}$ share the same connected components, each frontier has the same $\{d_{n_1,f} \}$ and $\{ d_{n_2,f} \}$. 
In addition, frontiers $f$ and $f'$ must be connected if they connect to at least one terminals (i.e., $t_{n_1,f}>0$ and $t_{n_2,f} >0$).
If (1) $\{ c_{n_1,f} \} = \{ c_{n_2,f} \}$ and (2) ($t_{n_1,f}= 0$ and $t_{n_2,f}=0$) or ($t_{n_1,f}> 0$ and $t_{n_2,f}>0$) for all $f$ in $\mathbb{F}_l$, nodes derived from $n_1$ and $n_2$ with the same states of edges $e_{l+1},\ldots, e_{|\mathbb{E}|}$ have the same attributes on the frontiers, and thus they make a transition to the same sink nodes.
 \hfill{} $\square$ 
 
 The probabilities of the merged nodes are aggregated to one node. The probabilities $p_c$ and $p_d$ are consistent, regardless of whether or not the nodes are merged.
 These procedures do not sacrifice the exactness of the network reliability.

\subsubsection{Deleting Procedure}
\label{sssec:delete}
The size of the S$^2$BDD increases exponentially as the size of the graph increases.
If the size of S$^2$BDD increases, the computation cost increases to obtain the lower and upper bounds of the network reliability because it takes a large time to construct the S$^2$BDD.
Hence, we control the size of S$^2$BDD by specifying the maximum size $w$.
The deleting procedure deletes the nodes so that the size of an S$^2$BDD is not larger than $w$. 
One of major difficulties in designing this procedure pertains to which nodes should be kept in the S$^2$BDD for achieving higher efficiency and accuracy. 
According to Theorems \ref{theo:sample1} and \ref{theo:sample2},  the number of samples effectively decreases as the probabilities $p_c$ and $p_d$ increase.
We identify intermediate graphs in which terminals are highly likely connected or disconnected after processing a small number of edges.
We make the following key observations in terms of the connectivity of terminals:
\begin{description}
\item[Observation 1] The terminals in the intermediate graph corresponding to node $n$ are highly likely connected if $t_{n,f}$ is large for $\exists f \in \mathbb{F}_l$.
\item[Observation 2] The terminals in the intermediate graph corresponding to node $n$ are highly likely disconnected  if $d_{n,f}$ is small and $t_{n,f}>0$ for $\exists f \in \mathbb{F}_l$.
\end{description}
Furthermore, if the probability of node $p_n$ is high and node $n$ makes a transition to sink nodes, $p_c$ and $p_d$ increase considerably.
Based on these observations, we define a heuristic function based on our observations.
We compute the priorities of nodes from their attributes by the heuristic function and preferentially keep high-priority nodes.
The heuristic function $h$ to compute the priority of node $n$ is as follows:

\begin{equation}{\textstyle
h(n) =   p_n \cdot \max_{f \in \mathbb{F}} \left( \frac{t_{n,f}}{k}, \frac{1}{d_{n,f}} \right) \mathrm{if}~t_{n,f} >0.}
\end{equation}
This function outputs larger value when (1) a frontier is connected to at least one terminals and (2) the frontier is connected to  a large number of terminals or (3) the frontier has a small number of uncertain edges.
In the former case, the terminals are likely connected, and in the latter case, the terminals are likely disconnected.
Low-priority nodes (i.e., $n$ with small $h(n)$) are then deleted from an S$^2$BDD.

\subsubsection{Sampling procedure}
\label{sssec:sampling}

Our approach samples possible graphs so that it avoids sampling the possible graphs that are used to compute the lower and upper bounds of network reliability, for satisfying the requirements of the stratified sampling.
We sample the possible graphs from the set of possible graphs that in which terminals are not c\-o\-n\-n\-e\-c\-t\-e\-d/d\-i\-s\-c\-o\-n\-n\-e\-c\-t\-e\-d yet.
We denote by $\mathbb{W}_u^{\mathcal{G}}$ such set of possible graphs, and the set is obtained from intermediate graphs corresponding to the deleted nodes and nodes in the S$^2$BDD.
We employ dynamic programming for efficiently sampling possible graphs from $\mathbb{W}_u^{\mathcal{G}}$.
In addition, we use the idea of the {\it stratified random sampling} \cite{thompson2002sampling} for determining the number of samples for subgroups that are  partial $\mathbb{W}_u^{\mathcal{G}}$.

We first divide $\mathbb{W}_u^{\mathcal{G}}$ into subgroups and then randomly sample possible graphs from each subgroup.
The number of samples for each subgroup is taken in proportion to the sum of the probabilities of the intermediate graphs in the subgroup.
We here explain only how to divide the deleted nodes and how to decide the number of samples for them.  As for the nodes in S$^2$BDD, each subgroup is the set of possible graphs obtained from the intermediate graph corresponding to the node, and the number of samples is computed from its probabilities. 

We divide the set of intermediate graphs for deleted nodes into subgroups according to original BDD layers instead of the node itself. 
This is because probabilities of deleted nodes are typically quite small to decide the number of samples.
$\mathbb{W}_u^{\mathcal{G}_l}$ and $s_l$ are the set of intermediate graphs corresponding to the deleted nodes at layer $l$ and the number of samples at layer $l$, respectively.
$s_l$ is computed by multiplying $s$ and the total probabilities $\hat{p_{s_l}}$ of deleted nodes at layer $l$.
We compute $\hat{p_{s_l}}$ from the attributes maintained by the S$^2$BDD by the following equation:
\begin{equation}\textstyle
\hat{p_{s_l}} =  1 - \sum_{i=1}^{l-1} p_{s_{i}} - p_{\mathbb{N}_{next}} - p_c - p_d,
\end{equation}
where $ p_{\mathbb{N}_{next}}$ denotes the sum of probabilities of $n \in \mathbb{N}_{next}$. 
$\hat{p_{s_l}}$  is the expected sum of probabilities of deleted nodes.
This is because $\hat{p_{s_l}}$ indicates the sum of probabilities in $\mathbb{N}_l$ when the number of nodes at layer $l+1$ reaches the maximum size.
The number of samples $s_l$ at layer $l$ becomes $s \cdot \hat{p_{s_l}}$.
The dynamic programming and stratified random sampling improve the efficiency of sampling while keeping the unbiased sampling. 

\subsection{Complexity}
We explain the time and space complexities of our approach.
\begin{theo}
Given the uncertain graph $\mathcal{G}$, the updated number of samples $s'$, and the maximum width of S$^2$BDD $w$, the time and space complexities of our approach are $O(w^2 \log  w + s'(|\mathbb{V}|+ |\mathbb{E}|))$ and $O(w\log w +|\mathbb{V}|+ |\mathbb{E}|)$, respectively.
\end{theo}
{\it Proof: }
The time complexity of our approach is divided into two parts; constructing S$^2$BDD and sampling.
To construct S$^2$BDD, our construction method compares attributes on each node each other for generating and merging procedures.
The number of attributes on each node increases in proportion to the number of frontiers.
The number of frontiers is $O(\log w)$ because the number of existent/non-existent edges is at most $\log w$.
Thus, the time complexity for constructing S$^2$BDD is $O(w^2 \log  w )$. 
The time complexity of sampling is $O(s'(|\mathbb{V}|+ |\mathbb{E}|))$.
Therefore, the time complexity of our approach is $O(w^2 \log  w + s'(|\mathbb{V}|+ |\mathbb{E}|))$.

The space complexity depends on the size of S$^2$BDD and the uncertain graphs. The size of S$^2$BDD is the number of nodes multiplied by the number of attributes on each node. Therefore, the space complexity is $O(w\log w +|\mathbb{V}|+ |\mathbb{E}|)$. 
\hfill{} $\square$


\section{Extension}
\label{ssec:graphreduction}

The computation cost of our approach depends on the size of the uncertain graphs as well as the number of samples.
The computation cost decreases as the size of the uncertain graphs decreases.
Therefore, we propose an extension technique to efficiently reduce the size of graphs while preserving the accuracy.
The extension technique preprocesses the uncertain graphs before sampling possible graphs and constructing an S$^2$BDD.
It not only improves the efficiency but also improves the accuracy of the approximation.
The extension technique uses {\it 2-edge-connected components} for reducting the size of uncertain graphs \cite{chang2013efficiently}.

\begin{defi}[2-edge-connected component]
Given a graph $G= (\mathbb{V}, \mathbb{E})$, an edge is called a {\it bridge} if $G$ is disconnected after the removal of the edge from $\mathbb{E}$.
Vertices that are connected by bridges are called {\it articulation points}.
A subgraph $C = (\mathbb{V}_C, \mathbb{E}_C)$ of $G$ is a 2-edge connected component if $C$ is still connected after the removal of any edges from $\mathbb{E}_C$.
We denote the sets of bridges, articulation points, and 2-edge connected components  by $\mathbb{B}$,  $\mathbb{A}$, and $\mathbb{C}$, respectively
\end{defi}

The 2-edge-connected components, bridges, and articulation points provide sets of edges (and vertices) such that the uncertain graph is disconnected or still connected when the edges (and vertices) are deleted.
Because we can compute 2-edge connected components only by using the network topology of a given uncertain graph, we precompute them as an index.

The extension technique consists of three phases; (1) pruning, (2) decomposing, and (3) transforming.
In the pruning phase, we first compute $\mathcal{G}'$ such that $R[\mathcal{G}] = R[\mathcal{G}']$. 
The number of edges in $\mathcal{G}'$ is smaller than that in $\mathcal{G}$ by pruning edges and vertices that do not affect computing the network reliability.
Next, in the decomposing phase, we compute the subgraphs $\mathcal{G}_1, \ldots, \mathcal{G}_m$ where $R[\mathcal{G}'] = \Pi_{i=1}^m R[\mathcal{G}_i]$.
Finally, in the transforming phase, we compute $\mathcal{G}_i'$ such that $R[\mathcal{G}_i] = R[\mathcal{G}_i']$ for all $1 \leq i \leq m$.
Since we transform the graph into a smaller graph, the number of edges in $\mathcal{G}_i'$ is smaller than that in $\mathcal{G}_i$.

{\bf Prune: }
We prune vertices and edges that do not affect the network reliability. 
A vertex (or an edge) is unnecessary if the graph is partitioned after the removal of the vertex (or edge) from $\mathcal{G}$ and one of the partitioned graphs does not include terminals.
A naive approach deletes each articulation point and bridge, and then checks whether partitioned graphs include terminals or not.
This approach incurs  $O((|\mathbb{B}|+|\mathbb{A}|)(|\mathbb{V}|+|\mathbb{E}|))$ time complexity.
To improve the efficiency, we reconstruct the uncertain graph based on the 2-edge connected components.
To do so, we first unite the set of vertices and edges included in $C \in \mathbb{C}$ to form a single vertex $v_c$. 
We then set every articulation point included in $C$ as vertex $v_a$ and set edges between $v_a$ and $v_c$.
The other vertices and edges that are not included in $\mathbb{C}$ are still in the reconstructed graphs.
Therefore, the vertices of the reconstructed graph indicate $\mathbb{C}$, $\mathbb{A}$, and the vertices that are not included in $\mathbb{C}$.  
If  any vertex in $C$ except for articulation points is a terminal, $v_c$ is also a terminal. 
The reconstructed graph is structured as a tree structure because the 2-edge connected components are connected to the other components by a single edge.
To compute the necessary vertices and edges, we compute the minimum Steiner tree for terminals in the reconstructed graph.
The minimum Steiner tree includes only the necessary vertices and edges to compute the network reliability because it includes only the edges and vertices that all the terminals are connected.
Its computation cost is $O(|\mathbb{V}|)$, because the minimum Steiner tree in a tree structure is computed by a depth first search from a terminal.

{\bf Decompose: } We decompose the graph because the time complexity for computing the network reliability on decomposed graphs becomes smaller than that on that original uncertain graph.
The decomposed graph has fewer edges than the original uncertain graph.
We decompose the graph according to the following lemma:

\begin{lemma}
Given an uncertain graph and a set of bridges, we obtain $R[\mathcal{G},\mathbb{T}] = p_b \cdot \prod_{i=1}^m R[\mathcal{G}_i,\mathbb{T}_i]$, where $p_b=\prod_{e_b \in \mathbb{B}} p(e_b)$ and $\mathbb{T}_i$ is the set of terminals for $\mathcal{G}_i$.
\end{lemma}
{\it Proof: } 
Given intermediate graph $\mathcal{G}_{\mathbb{E}}(\mathbb{E}_{\exists}, \mathbb{E}_{\lnot})$ and edge $e \in \mathbb{E}\backslash (\mathbb{E}_{\exists} \cup \mathbb{E}_{\lnot})$, the network reliability is computed using the Factoring Theorem \cite{colbourn1987combinatorics}:
\begin{eqnarray}\label{eq:bridge}
R[\mathcal{G}_{\mathbb{E}}(\mathbb{E}_{\exists}, \mathbb{E}_{\lnot})] &=& p(e) \cdot R[\mathcal{G}_{\mathbb{E}}(\mathbb{E}_{\exists} \cup e, \mathbb{E}_{\lnot})] \nonumber \\
& &+ (1-p(e)) \cdot R[\mathcal{G}_{\mathbb{E}}(\mathbb{E}_{\exists}, \mathbb{E}_{\lnot} \cup e)].
\end{eqnarray}
If we select bridge $e_b =(v,v') \in \mathbb{B}$ as $e$ in Equation (\ref{eq:bridge}), $R[\mathcal{G}_{\mathbb{E}}(\mathbb{E}_{\exists}, \mathbb{E}_{\lnot} \cup e)]$ is zero because terminals in $\mathcal{G}_{\mathbb{E}}(\mathbb{E}_{\exists}, \mathbb{E}_{\lnot} \cup e)$ are disconnected.
Therefore, we obtain the following equation:
\begin{equation}
R[\mathcal{G}_{\mathbb{E}}(\mathbb{E}_{\exists}, \mathbb{E}_{\lnot})] = p(e_b) \cdot R[\mathcal{G}_{\mathbb{E}}(\mathbb{E}_{\exists} \cup e_b, \mathbb{E}_{\lnot})].
\end{equation}
For connecting all the terminals, $e_b$ must be existent, and thus we can decompose  the intermediate graph $\mathcal{G}_{\mathbb{E}}$ into two graphs $\mathcal{G}_{\mathbb{E}_1}$ and $\mathcal{G}_{\mathbb{E}_2}$.
We also divide the terminals $\mathbb{T}$ into $\mathbb{T}_1$ and $\mathbb{T}_2$ for $\mathcal{G}_{\mathbb{E}_1}$ and $\mathcal{G}_{\mathbb{E}_2}$, respectively; $\mathbb{T}_1$ includes $\{t \in \mathbb{T}, v, v' | t, v,v' \in \mathbb{V}_1 \}$ (similarly, $\mathbb{T}_2$).
Thus, $R[\mathcal{G}_{\mathbb{E}}] = p(e_b) \cdot R[\mathcal{G}_{\mathbb{E}_1}] R[\mathcal{G}_{\mathbb{E}_2}]$.
$\mathcal{G}_{\mathbb{E}_1}$ and $\mathcal{G}_{\mathbb{E}_2}$ are decomposed in the same manner.
Then, we obtain $R[\mathcal{G}] = p_b \cdot \prod_{i=1}^m R[\mathcal{G}_i,\mathbb{T}_i]$.
\hfill{} $\square$
\vspace{1mm}

We decompose the uncertain graph into several subgraphs based on the above lemma. Its computation cost is $O(|\mathbb{B}||\mathbb{V}|)$ because we check whether decomposed graphs include terminals or not for each bridge.

{\bf Transform: } We transform the graph to reduce its size.
We delete and add the following edges and vertices without sacrificing the exactness of the network reliability:
\begin{itemize}
\item Sequential edges ($e = (v, v'), e' =(v, v'')$): Delete $v$, $e$ and $e'$, and add a new edge with probability $p(e) \cdot p(e')$ between $v'$ and $v''$, provided that $v$ is not a terminal and its degree is two. 
\item Parallel edges ($e =(v, v'), e' = (v, v')$): Delete $e$ and $e'$, and add a new edge with probability $(1-(1-p(e)\cdot(1-p(e'))$ between $v$ and $v'$.
\item Loop : Delete the loop because loops do not contribute to the network reliability. Note that transforming sequential and parallel edges can generate loops.
\end{itemize}
We iteratively repeat this process until the graph does not change.
The computation cost is $O(\gamma \cdot |\mathbb{V}|\cdot {d_{avg}}^2)$ where $\gamma$ and $d_{avg}$ are the number of repetitions and the average degree of the vertices, respectively.
\begin{algorithm}[!t] 
	{\footnotesize
	\caption{Computing the approximate network reliability}	\label{alg:whole}
		\DontPrintSemicolon
			    \SetKwInOut{Input}{input}
	            \SetKwInOut{Output}{output}
	             \SetKwFunction{Preprocess}{Preprocess}
	             \SetKwFunction{Construction}{Construction}	             
	             \SetKwFunction{Sampling}{Sampling}
	            \SetKwFunction{update}{update}
	            \Input{Uncertain graph $\mathcal{G}$, terminals $\mathbb{T}$, maximum BDD size $w$, size of samples $s$,  2-edge connected components $\mathbb{C}$, bridges $\mathbb{B}$, articulation points $\mathbb{A}$}
	            \Output{Approximate network reliability $\hat{R}$}
            	{\bf procedure} our approach\\	
            	set $\mathbb{T}$ to $\mathcal{G}$;\\
            	$\hat{R}, \mathcal{S}_\mathcal{G} \leftarrow$ \Preprocess{$\mathcal{G}$, $\mathbb{T}$, $\mathbb{C}$, $\mathbb{B}$, $\mathbb{A}$};\\
            	\For{$\mathcal{G}_i \in \mathcal{S}_\mathcal{G}$}{
	            	$r \leftarrow$ \Construction{$\mathcal{G}_i$, $w$, $s$ };\\
	            	$\hat{R} \leftarrow \hat{R} \cdot r$;\\
            	}
              {\bf return} $\hat{R}$;\\
              {\bf end procedure}
          }
\end{algorithm}
\begin{algorithm}[!t] 
	{\footnotesize
	\caption{Constructing S$^2$BDD}	\label{alg:construct}
		\DontPrintSemicolon
			    \SetKwInOut{Input}{input}
	            \SetKwInOut{Output}{output}
	             \SetKwFunction{Construction}{Construction}
	             \SetKwFunction{Sampling}{Sampling}
	            \SetKwFunction{set}{set}
	            \Input{Uncertain graph $\mathcal{G}$, maximum size $w$, number of samples $s$}
	            \Output{Approximate network reliability $\hat{R}$}
            	{\bf procedure} \Construction{$\mathcal{G}$, $w$, $s$}\\	
	            Ordering($\mathbb{E}$);\\ 
	            $p_c, p_d, \hat{p_{s_l}}, c \leftarrow 0$;\tcc*[r]{initialize probabilities and sampling count}
	            $s' \leftarrow s$;\\
	            $\mathbb{N} \leftarrow$ CreateRoot; $\mathbb{F} \leftarrow null$; \\
            	\For{$l$ for $1, \ldots , |\mathbb{E}|$}{
            	$p_{\mathbb{N}}, p_{s_i} \leftarrow 0$;\\
            	$\mathbb{F}' \leftarrow \mathbb{F}$; compute $\mathbb{F}$ based on $e_l$;\\
	            	 \While{$\mathbb{N}$ is empty}{
						  $n \leftarrow \mathbb{N}.pop$;\\
						  \For{$\mathit{state} \in \{$ $\mathit{non}$-$\mathit{existent}$, $\mathit{existent}\}$}{
							  \set{$n$, $\mathbb{F}'$, $\mathbb{F}$, $\mathit{state}$, $\mathcal{G}$, $e_l$};\\		
							  {\bf if} $n$ {\it is} $0${\it -sink} {\bf then} $p_d \leftarrow p_d+ p_n$;\\
							  {\bf else if} $n$ {\it is} $1${\it -sink} {\bf then} $p_c \leftarrow p_c+ p_n$;\\
	 						  \Else{
		 						  \If{$\mathit{hashmap}[n]$ is not null}{
			 						  $p_{\mathit{hashmap}[n]} \leftarrow p_{\mathit{hashmap}[n]} + p_n$;
		 						  }
								  \Else{
									  \If{$|\mathbb{N}_{\mathit{next}}|$ $\leq w$}{
										  $h_n$ $\leftarrow$ $h(n)$;\\									
										  $\mathbb{N}_{\mathit{next}}$.add($n$);
										  $\mathit{hashmap}[n] \leftarrow n$;
										  $p_{\mathbb{N}_{next}} \leftarrow p_{\mathbb{N}_{next}} + p_n$;
									  }
									  \Else{
									     $p_{s_i} \leftarrow p_{s_i} + p_n$;\\
									     \For{ $i$ for $1, \ldots , \lfloor s' \cdot (1-\hat{p_{s_l}}-p_{\mathbb{N}_{\mathit{next}}}-p_c-p_d) \rfloor$}{ 
									     {\bf if} \Sampling{$\mathcal{G}$, $n$} {\bf then} $c\leftarrow c+1$;}
									  }
								  }
							  }
						  }
					   \If{$c + \lfloor s' \cdot p_{\mathbb{N}_{\mathit{next}}} \rfloor \geq s'$}{
						   \For{$n \in \mathbb{N}$}{
						     \For{ $i$ for $1, \ldots , \lfloor s'\cdot p_{\mathbb{N}_{\mathit{next}}} \rfloor$}{ 
						   			{\bf if} \Sampling{$\mathcal{G}$, $n$} {\bf then} $c\leftarrow c+1$;}
							 } 
					   {\bf break};		
					  }
					  \If{$\mathbb{N}_n$ is empty}{
						  {\bf break};	
					  }					 
					  $\mathbb{N} \leftarrow \mathbb{N}_{\mathit{next}}$;\\
					  sort $\mathbb{N}$ in descending order of $h(n)$;\\
					  $\hat{p_{s_l}} \leftarrow \hat{p_{s_l}}+p_{s_i}$;
					  compute $s'$;
  					  clear $\mathbb{N}_{\mathit{next}}$;
 					  clear $\mathit{hashmap}$;\\		
  					  }			            
  					  }
               compute $\hat{R}$ based on the sampling;\\
              {\bf return} $\hat{R}$;\\
              {\bf end procedure}
          }
\end{algorithm}

Consequently, the extension technique effectively reduces the computation cost for computing the network reliability with a small preprocessing time.
Furthermore, it improves the accuracy of the sampling technique.

\begin{algorithm}[t]
{\footnotesize
	\caption{Extension technique}	\label{alg:graphreduction}
		\DontPrintSemicolon
	            \SetKwInOut{Input}{input}
	            \SetKwInOut{Output}{output}
	            \SetKwFunction{Reconstruct}{Reconstruct}
	             \SetKwFunction{Preprocess}{Preprocess}
	            \Input{Uncertain graph $\mathcal{G}$, terminals $\mathbb{T}$, 2-edge connected components $\mathbb{C}$, bridges $\mathbb{B}$, articulation points $\mathbb{A}$}
	            \Output{Probability $p_b$, the set of decomposed graphs $\mathcal{S}_{\mathcal{G}}$}
            	{\bf procedure} \Preprocess{$\mathcal{G}$, $\mathbb{C}$, $\mathbb{B}$, $\mathbb{A}$}\\	
            	\tcc{Prune}
            	$\mathcal{G}_r \leftarrow$ \Reconstruct{$\mathcal{G}$};\\
            	Compute the minimum Steiner tree $\mathcal{T}$ for $\mathcal{G}_r$ and terminals;\\
            	Delete edges and vertices of $\mathcal{G}$ not included in $\mathcal{T}$;\\
            	\tcc{Decompose}
            	$p_b \leftarrow \prod_{e_b \in \mathbb{B}}~p(e_b)$;\\
            	Delete the set of bridges in $\mathcal{G}$;\\
            	$\mathcal{S}_{\mathcal{G}} \leftarrow$ the set of disconnected graphs;\\
            	\tcc{Transform}
            	\For{$\mathcal{G}' \in \mathcal{S}_{\mathcal{G}}$}{
		          	\While{1}{
			          	\For{$v \in$ $\mathbb{V}$ of $\mathcal{G}'$}{
				          	\If{$v$ connects to edge $e =(v,v)$}{delete $e =(v,v)$;}
				          	\If{$v \notin \mathbb{T}$ and $v$ connects to just two edges $e=(v,v')$ and $e'=(v,v'')$}{
					          	delete $e$ and $e'$ from $\mathcal{G}'$;\\ 
					          	add a new edge $(v', v'')$ with probability $p(e)\cdot p(e')$;\\
				          	}
			          	}
			          	\For{$v \in$ $\mathbb{V}$ of $\mathcal{G}'$}{
				          	\For{$\forall$ pair of $u$ and $u'$ $\in$ the set of neighbor vertices of $v$}{
					          	\If{$u = u'$}{
						          	delete edge $e =(v,u)$ and $e'=(v,u')$;\\
						          	add a new edge $(v, u)$ with probability  $(1-(1-p(e)\cdot(1-p(e'))$;\\
						          	}
				          	}
			          	}
			          	\If{The number of edges does not change}{\bf{break};}
				   }
			   }
              {\bf return} $p_b$, $\mathcal{S}_{\mathcal{G}}$;\\
              {\bf end procedure}
              }
\end{algorithm}

\begin{theo}
Given $\mathcal{G}_1, \ldots, \mathcal{G}_m$ such that $R[\mathcal{G}] = p_b \cdot \Pi_{i=1}^m R[\mathcal{G}_i]$, the variance of the network reliability decreases for $0 <\hat{R} < 1$ and $0 < p_b < 1$. 
\end{theo}
{\it Proof: } The network reliability is denoted by $\hat{R} = p_b \cdot \Pi_{i=1}^m \hat{R}[\mathcal{G}_i]$. The valiance is computed as follows:

\begin{eqnarray}\textstyle
Var[\hat{R}] &=& Var[p_b \cdot \Pi_{i=1}^m \hat{R}[\mathcal{G}_i]] \nonumber \\
&=& (Var[p_b]+{p_b}^2)(Var[\hat{R}[\mathcal{G}_1]]+{\hat{R}[\mathcal{G}_1]}^2)\cdots \nonumber \\
& & (Var[\hat{R}[\mathcal{G}_m]]+\hat{R}[\mathcal{G}_m]^2) - {p_b}^2 \cdot \Pi_{i=1}^m \hat{R}[\mathcal{G}_i]^2 \nonumber \\
&=&\textstyle {p_b}^2\Pi_{i=1}^m (Var[\hat{R}[\mathcal{G}_i]] + {\hat{R}[\mathcal{G}_i]}^2) - {p_b}^2 \Pi_{i=1}^m {\hat{R}[\mathcal{G}_i]}^{2} \nonumber \\
&=&\textstyle {p_b}^2\Pi_{i=1}^m \left( \frac{\hat{R}[\mathcal{G}_i](1-\hat{R}[\mathcal{G}_i])}{s} + {\hat{R}[\mathcal{G}_i]}^2 \right) - {p_b}^2\Pi_{i=1}^m {\hat{R}[\mathcal{G}_i]} ^{2}\nonumber \\
&=&\textstyle{p_b}^2\Pi_{i=1}^m \hat{R}[\mathcal{G}_i] \left( \frac{(1 + (s-1) \hat{R}[\mathcal{G}_i])}{s} \right) - {p_b}^2\Pi_{i=1}^m {\hat{R}[\mathcal{G}_i]}^{2}\nonumber \\
&<&\textstyle \frac{{p_b}^2\Pi_{i=1}^m \hat{R}[\mathcal{G}_i]}{s} - \frac{{p_b}^2\Pi_{i=1}^m {\hat{R}[\mathcal{G}_i]}^2}{s} \nonumber \\
&=&\textstyle {p_b} \frac{\hat{R} (1-\hat{R})}{s} < \frac{\hat{R}(1-\hat{R})}{s} 
\end{eqnarray}
Note that $Var[p_b]=0$. $Var[\hat{R}]$ is smaller than the variance of the network reliability of the original graph. \hfill{} $\square$

\section{Algorithm of our approach}
\label{sec:algo}

 \begin{table*}[ttt]
 \begin{center}
 \caption{Dataset}	\label{dataset}
 \vspace{-3mm}
 \begin{tabular}{|l|l|l|l|l|l|l|} \hline
 Name & Abbr. &Type &$\#$vertices & $\#$edges  & Avg.\ Deg & Avg.\ Prob  \\ \hline\hline
  Zachary-karate-club & Karate  &Social & 34 & 78 & 4.59 & 0.527 \\
  American-Revolution& Am-Rv &Affiliation & 141& 160 & 2.27 & 0.528	 \\
 DBLP before 2000& DBLP1 &Coauthorship & 25,871 & 108,459 & 8.38 & 0.222\\ 
 DBLP after 2000& DBLP2	&Coauthorship& 48,938 & 136,034 & 5.56 & 0.203 \\
 Tokyo & Tokyo& Road network &26,370 & 32,298 & 2.45 & 0.391 \\
 New York City& NYC& Road network &180,188 & 208,441 & 2.31 & 0.294\\
 Hit-direct& Hit-d& Protein &18,256 & 248,770 & 27.25 & 0.470 \\
	 \hline
 \end{tabular}
 \end{center}
 \vspace{-2mm}
 \end{table*}
In this section, we explain the entire algorithm of our approach.
Algorithm \ref{alg:whole} shows the pseudo-codes.  
Our approach first preprocesses uncertain graphs and obtains decomposed uncertain graphs (line 3).
For each decomposed graph, it then constructs an S$^2$BDD to compute  the approximate network reliability of the decomposed graphs (lines 4--5).
The product of the network reliability of each decomposed graph is the original network reliability (line 6).

Algorithm \ref{alg:construct} shows the pseudo-codes for  the construction of an S$^2$BDD.  
We process edges in a predefined order, and compute the set of frontiers (lines 6--8).
For each node at  layer $l$, we compute the nodes at layer  $l+1$ according to the states of the edges (lines 11--12).
The {\sf set} function (line 12) sets attributes on the new node to $n$ and checks whether the terminals are connected or disconnected based on Lemmas 1 and 2. 
If the new node are 0-sink and 1-sink, we add $p_n$ to $p_d$ and $p_c$, respectively (line 13--14). 
Otherwise, we compute hash values for $n$, and if the hash of $n$ is not null, we add the probability $p_n$ to the node in the hash (lines 16--17). 
If the hash is null with respect to $n$, it inserts $n$ into the set $\mathbb{N}_{next}$ of nodes at layer $l+1$ and into the hash after computing their priorities (lines 19--21).
If the number of nodes in $\mathbb{N}_n$ exceeds the maximum size $w$, we delete $n$ and pick possible graphs as samples from $n$ (lines 22--25).
After sampling an enough number of possible graphs, we sample form the nodes in the S$^2$BDD (lines 26--29).

Algorithm \ref{alg:graphreduction} shows the pseudo-codes for the extension technique.  
The extension technique  first reconstructs the uncertain graph (line 2).
Then, it computes the minimum Steiner tree for the reconstructed graph and prunes the edges and vertices  that are not included in the Steiner tree from the original uncertain graph (lines 3--4).
To decompose the graph, we compute the product of the probabilities of bridges $p_b$ (line 5).
Then, we delete bridges from the uncertain graph, and the disconnected subgraphs are inserted into the set of decomposed uncertain graphs (lines 6--7).
For each decomposed graph, it transforms vertices and edges that satisfy the transformation rules (lines 8--20).

\section{Experiment}
\label{sec:experiment}
We evaluate our approach in terms of efficiency, accuracy, and memory usage.

\subsection{Dataset}

We summarize the datasets in Table \ref{dataset}.
The first two datasets; Zachary-karate-club and American-revolution are small datasets for evaluating accuracy, which are extracted from KONECT\footnote{http://konect.uni-koblenz.de/}.
We randomly assign probabilities based on the uniform distribution \cite{cheng2016distr}. 
The other five datasets; DBLP before 2000, DBLP after 2000, Tokyo, New York City, and Hit-direct, are large datasets.
Edge existence probabilities for each large dataset are assigned based on the attributes of the edges in each dataset.
DBLP before 2000 and DBLP after 2000 are graphs extracted from DBLP\footnote{http://dblp.uni-trier.de/}, where vertices and edges are authors and co-author, respectively.
We compute the edge existence probabilities by $\frac{\log(\alpha + 1)}{\log(\alpha_M + 2)}$, where $\alpha$ and $\alpha_M$ denote the number of co-authors and the maximum in each dataset, respectively~ \cite{ceccarello2017clustering}. 
The Tokyo and New York City datasets are road networks extracted from OpenStreetMap\footnote{https://www.openstreetmap.org}.
We compute the edge existence probabilities in the same manner as with the DBLP datasets, although we use road lengths instead of the number of co-authors.
Note that both the Tokyo and New York City datasets are not planar graphs.
Hit-direct is a protein-protein interaction network extracted from the Human Genome Center\footnote{http://hintdb.hgc.jp/htp/download.html.}.
We use the interaction scores $\in (0,1]$ of interactions as  the edge existence probabilities.

\begin{figure*}[ttt]
  \centering
   \subfloat[$k=5$]{\epsfig{file=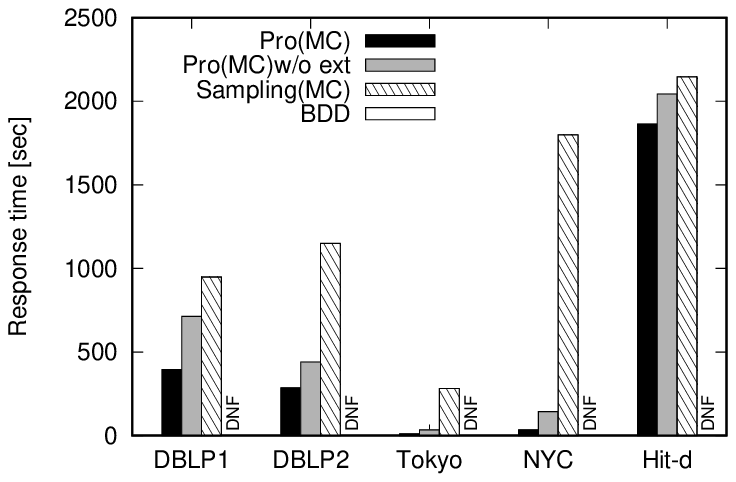,width=0.33\linewidth}
   \label{fig:cpu1}}
   \subfloat[$k=10$]{\epsfig{file=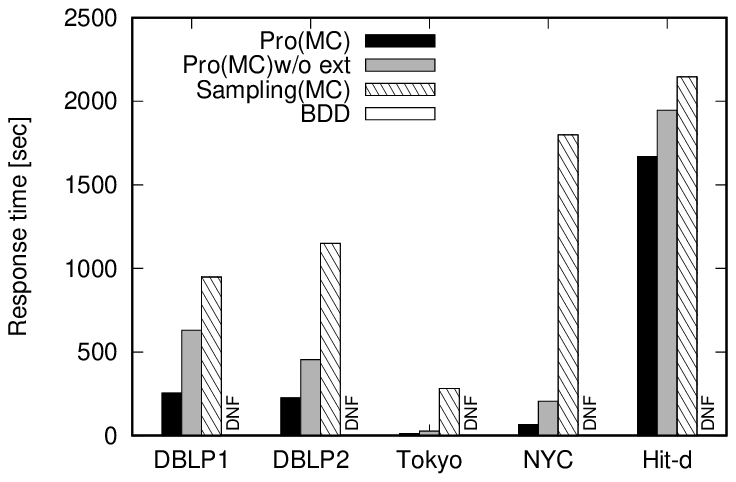,width=0.33\linewidth}
   \label{fig:cpu1}}
    \subfloat[$k=20$]{\epsfig{file=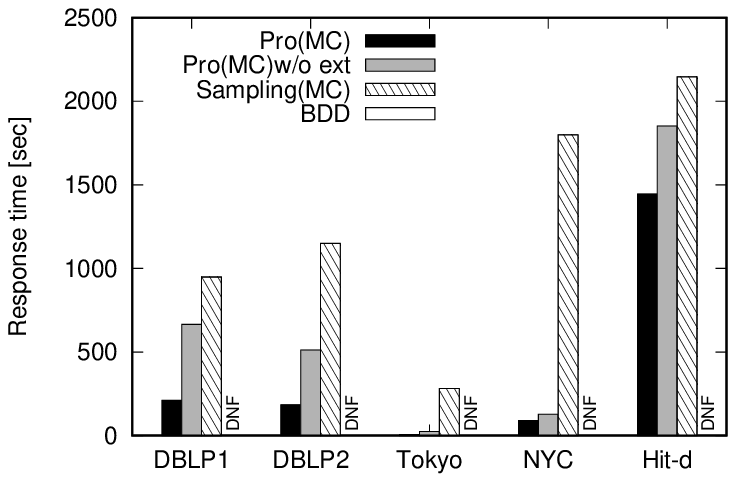,width=0.33\linewidth}
    \label{fig:cpu1}}
    \vspace{-3mm}
   \caption{Overview of efficiency}
   \label{fig:efficiency}
   \vspace{-5mm}
   \end{figure*}

  \begin{figure*}[ttt]
     \begin{minipage}[]{.49\textwidth}
  \centering
   \subfloat[Response time]{\epsfig{file=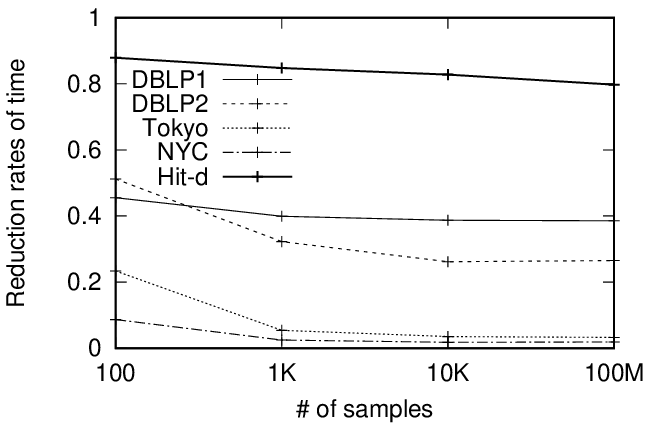,width=0.5\linewidth}
   \label{fig:cpu1}}
   \subfloat[$\#$ of samples]{\epsfig{file=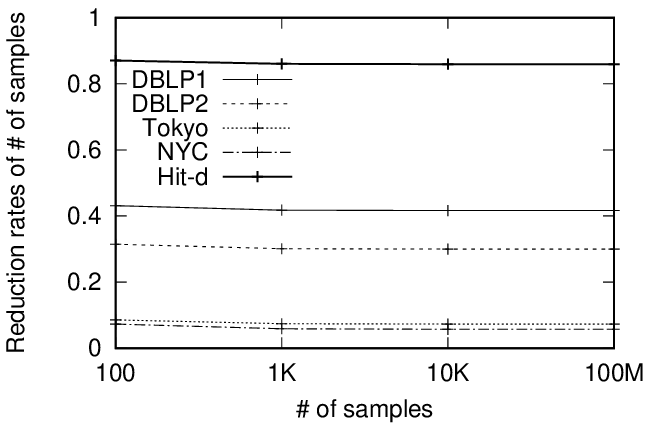,width=0.5\linewidth}
   \label{fig:cpu1}}
   \vspace{-4mm}
   \caption{Efficiency with varying the number of samples}
   \label{fig:speedupsample}
   \end{minipage}
    \hfil 
    \begin{minipage}[]{.49\textwidth}
    \centering
   \subfloat[Memory usage]{\epsfig{file=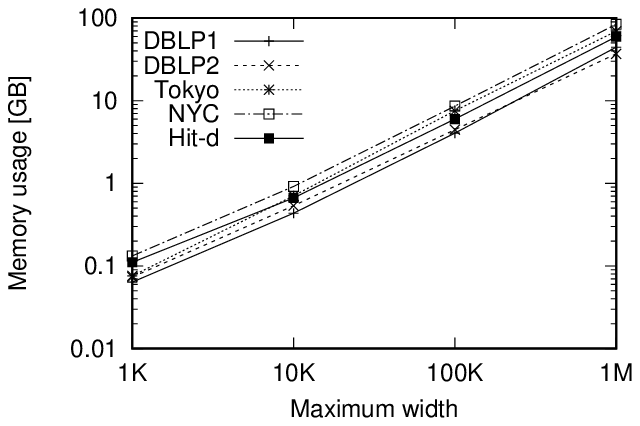,width=0.5\linewidth}
   \label{fig:cpu1}}
   \subfloat[Response time]{\epsfig{file=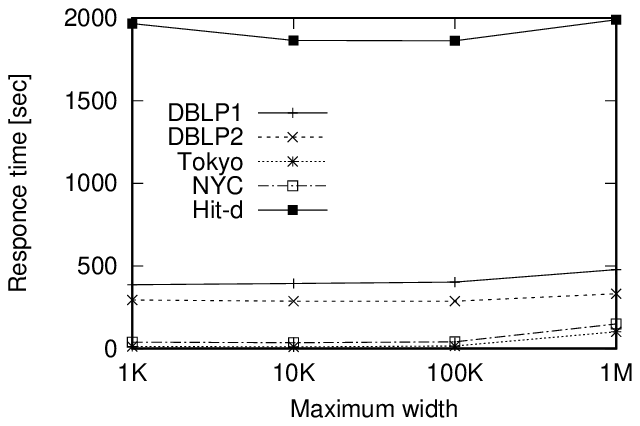,width=0.5\linewidth}
   \label{fig:cpu1}}
   \vspace{-4mm}
   \caption{Efficiency with varying the maximum width}
   \label{fig:maximumwidth}
    \end{minipage}
   \end{figure*}

\subsection{Setting and Implementation}

For each dataset, we generate 20 searches (except when we evaluate the accuracy, for which see Section \ref{ssec:accuracy}).
The terminals are selected randomly from vertices.
We vary the number of terminals $k$, the number of  samples $s$, and the maximum size  of the S$^2$BDD $w$.

Because the existence probabilities of possible graphs can be very small, we use the Boost.Multiprecision library, with precision of 10,000 decimal points, for the large datasets.
We compute the 2-edge-connected components using code provided by the authors \cite{chang2013efficiently}.
We compare our approach with two existing approaches; the sampling-based and BDD-based approaches.
The BDD-based approach uses the state-of-the art library, TdZDD.
All algorithms are implemented in C++, and run on a server with an Intel Xenon E7-8860v4 at 2.20GHz with 256GB RAM.

\subsection{Efficiency}
\label{ssec:efficiency}
We compare the efficiency of our approach with that of sampling-based and BDD-based approaches.  
Figure \ref{fig:efficiency} shows the response time for each large dataset when the numbers of terminals $k$ is set to 5, 10, and 20.
DNF indicates that we cannot compute the network reliability due to the lack of memory space.
We use Monte Carlo estimator for our approach and the sampling-based approach (denoted by {\sf Pro(MC)} and {\sf Sampling(MC)}, respectively) and set $s$ to 10,000.
For our approach, we set $w$ to 10,000.
We also evaluate our approach without the extension technique denoted by {\sf Pro(MC)w/o ext}.
We here omit the results of Horvitz-Thompson estimator because they are almost equivalent to those of Monte Carlo estimator.

The results show that our approach is more efficient than both of the sampling-based and the BDD-based approaches for all $k$.
The BDD-based approach cannot compute the network reliability because it runs out of memory.
Our approach achieves higher efficiency than the sampling-based approach because it reduces the number of samples.
Furthermore, we can see that the extension technique improves the efficiency.
In particular, our approach works well on the Tokyo and NYC datasets.
This is because the S$^2$BDD works well for planar-like graphs (even when they are not strictly planar graphs). 
In the Hit-direct dataset, the lower and upper bounds do not effectively become tight because the number of degrees is large.
Nevertheless, our approach is more efficient than the sampling-based approach.

\subsection{Effect of Number of Samples}
We evaluate the effect of the given number of samples.
Figure \ref{fig:speedupsample} shows (a) the rate of response time of our approach over that of the sampling-based approach and (b) the rate of updated samples $s'$ over $s$, varying the number of samples.
This figure shows that our approach becomes more efficient as the given number of samples increases.
This is because the reduction of the number of samples is more effective when the given number of samples is large.
Therefore, our approach more effectively works when we need a high accurate network reliability.

\subsection{Effect of Maximum Width}
\label{ssec:adaptivity}
We evaluate the effect of the given maximum width of S$^2$BDD.
The maximum width $w$ affects the memory usage and efficiency.
Figure \ref{fig:maximumwidth} shows (a) the memory usage and (b) the response time.
From Figure \ref{fig:maximumwidth}(a), we can see that the memory usage increases as the maximum width increases.
The memory usage depends on the maximum width but not depends on the size of graphs.
Our approach can be used for large-scale graphs in terms of memory usage.
From Figure \ref{fig:maximumwidth}(b), we can see that the response time does not largely depend on the maximum width.
When the maximum width is large, our approach can reduce the number of samples but takes a large computation cost for constructing S$^2$BDD.
Our approach is robust enough to the maximum width in terms of efficiency.
Consequently, our approach effectively decreases the response time even for large-scale graphs.


  \begin{table*}[t]
  \begin{minipage}[]{.33\textwidth}
  \begin{center}
  \caption{Accuracy on Karate dataset}
  \vspace{-3mm}
  \label{table:accuracykarate}
  \small
  \begin{tabular}{|l|l|l|l|} \hline
  \multirow{1}{*}{$k$}		&	\multirow{1}{*}{Method}		& 	\multicolumn{1}{|c|}{Variance}		&\multicolumn{1}{|c|}{Error rate} 							  \\\hline\hline
  \multirow{4}{*}{5}		& {\sf Pro(MC)} 		& 0.025		& 0.036	\\
									   & {\sf Pro(HT)} 		& 0.025		& 0.036	\\
  										&	{\sf Sampling(MC)}	& 0.025	 	&	0.037	\\													
  										&	{\sf Sampling(HT)}					& 0.029		& 	0.042	\\			
  														\hline						
  \multirow{4}{*}{10}		& {\sf Pro(MC)}		& 0.013 		& 	0.058	\\
										  & {\sf Pro(HT)}		& 0.014 		& 	0.059	\\
  										&	{\sf Sampling(MC)}	& 0.013 		& 0.058	\\									
  										&	{\sf Sampling(HT)}					&	0.015		& 0.062	\\
  												\hline						
  \multirow{4}{*}{20}		& {\sf Pro(MC)}	& 0.76 $\cdot 10^{-3}$	& 0.054	\\
										  & {\sf Pro(HT)}		& 0.85$\cdot 10^{-3}$ 	& 0.057	\\
  										&	{\sf Sampling(MC)}	&	0.78$\cdot 10^{-3}$  	& 0.056	\\							
  										&	{\sf Sampling(HT)}				&	0.86$\cdot 10^{-3}$ 	& 0.057	\\
  											\hline			
  \end{tabular}
  \end{center}
  \end{minipage}
  \hfil
  \begin{minipage}[]{.33\textwidth}
  \begin{center}
  \caption{Accuracy on Am-Rv dataset}
  \vspace{-3mm}
  \label{table:accuracyrevolution}
  \small
  \begin{tabular}{|l|l|l|l|} \hline
  \multirow{1}{*}{$k$}		&	\multirow{1}{*}{Method}		& 	\multicolumn{1}{|c|}{Variance}		&\multicolumn{1}{|c|}{Error rate} 							  \\\hline\hline
  \multirow{4}{*}{5}		& {\sf Pro(MC)}		& 0						& 0						\\
										  & {\sf Pro(HT)}	& 0						& 0						\\
  										&	{\sf Sampling(MC)}	& 0.43$\cdot 10^{-4}$	&	0.061	\\													
  										&	{\sf Sampling(HT)}	 				& 0.31$\cdot 10^{-4}$	& 	0.059	\\			
  														\hline						
  \multirow{4}{*}{10}		& {\sf Pro(MC)}		& 0 						& 	0						\\
									  & {\sf Pro(HT)}		& 0						& 0						\\
  										&	{\sf Sampling(MC)}	& 0.099$\cdot 10^{-5}$ 	& 0.38	\\									
  										&	{\sf Sampling(HT)}					&	0.12$\cdot 10^{-5}$	& 0.37	\\
  												\hline						
  \multirow{4}{*}{20}		& {\sf Pro(MC)} 		& 0 						& 0						\\
										  & {\sf Pro(HT)}		& 0						& 0						\\
  										&	{\sf Sampling(MC)}	&	0.10$\cdot 10^{-3}$  	& 1.00 	\\							
  										&	{\sf Sampling(HT)}					&	0.10$\cdot 10^{-3}$ 	& 1.00	\\
  											\hline			
  \end{tabular}
  \end{center}
  \end{minipage}
  \hfil
   \begin{minipage}[]{.33\textwidth}
\begin{center}
\small
\caption{Effect of extension technique}	\label{table:effectgraph}
\vspace{-3mm}
\begin{tabular}{|l|l|l|} \hline
\multirow{2}{*}{Dataset}	 			& Process time  &Reduced\\ 
													 &  [sec]  						 & graph size \\\hline\hline
Karate 								& 	0.0277$\cdot 10^{-3}$ 	& 0.757  \\   
Am-Rv 							&  0.310$\cdot 10^{-3}$		& 0.120  \\	 
DBLP1					& 0.060 	 	&	0.946 \\ 
DBLP2						& 1.61  		& 0.797 \\
Tokyo 						& 0.015  		& 0.425 \\
NYC						&	0.370   	&	0.279  \\
Hit-d				& 0.184   	&	0.982\\ \hline
\end{tabular}
\end{center}
\end{minipage}
\end{table*}

\subsection{Accuracy}
\label{ssec:accuracy}
We evaluate the accuracy of our approach compared with the sampling-based approaches.
For both approaches, we use Horvits-Thompson estimator (denoted by {\sf Pro(HT)} and {\sf Sampling(HT)}) as well as  Monte Carlo estimator.
Since the network reliability problem is $\#P$-complete, we cannot compute the exact answer for large datasets in terms of both response time and memory usage.
We use the Karate and Am-Rv datasets which can be computed the exact network reliability.
We evaluate the variance and the error rate to determine the accuracy of the approximation as follows:
{\it variance} $=\frac{\Sigma_{i=1}^{q_1} \Sigma_{j=1}^{q_2} (R_i -\hat{R}_{i,j})^2}{q_1 \cdot q_2}$ and
{\it error~rate} $=\frac{\Sigma_{i=1}^{q_1} \Sigma_{j=1}^{q_2} |R_i - \hat{R}_{i,j}|}{q_1 \cdot q_2 \cdot R_i}$,
where $R_i$ and $\hat{R}_{i,j}$ denote the $i$-th exact network reliability and the $j$-th approximate network reliability for the $i$-th search, respectively.
We generate 100 searches and compute the network reliability 100 times for each search (i.e., both $q_1$ and $q_2$ are 100).

Tables \ref{table:accuracykarate} and \ref{table:accuracyrevolution} show the accuracy on the Karate and Am-Rv datasets, respectively.
Table \ref{table:accuracykarate} shows that our approach outperforms the sampling-based approaches in terms of both of  the variance and error rate.
Comparing the variance between the estimators, the Monte Carlo estimator is slightly better than the Horvits-Thompson sampling.
This is because we sample possible graphs with replacement, and thus the Horvits-Thompson estimator is less effective.
Table \ref{table:accuracyrevolution} shows that  our approach always computes the exact network reliability on the Am-Rv dataset--- its error rate is zero.
Both of the existing sampling-based approaches have high error rates when $k=20$ although their variances are small.
Because the network reliability is very small, the sampling-based approaches rarely sample the possible graphs in which terminals are connected. 
Thus, the approximate network reliability is often zero, and the error rates are close to one.
From these results, we conclude that our approach can achieve less variance and error rate with fewer samples than the other approaches and compute the exact answer for small-scale graphs.

\subsection{Effect of Extension Technique}
Finally, we evaluate the performance of the extension technique.
The effect of the extension technique is detailed in Table~\ref{table:effectgraph} which shows the process time and the ratio of the maximum number of edges in decomposed graphs over the number of edges in the original uncertain graph.
The results show that the extension technique requires a very small time compared with computing the network reliability. 
Thus, it effectively reduces the total response time.
Since it reduces the size of uncertain graphs, it mitigates the computation cost for the S$^2$BDD.
The extension technique is effective for improving the efficiency of our approach.

\section{Conclusion}
\label{sec:conclusion}
In this paper, we proposed an efficient sampling-based approach for computing the approximate network reliability.
Our approach reduces the number of samples by using lower and upper bounds of the network reliability based on the stratified sampling.
We developed scalable and sampling BDD, called S$^2$BDD, which efficiently computes the bounds.
The S$^2$BDD preferentially searches for the possible graphs that highly improve the bounds.
We further developed the extension technique of our approach to reduce the size of graphs.
Experiments demonstrated that our approach is up to 51.2 times faster than the sampling-based approach with a higher accuracy.



\section*{Acknowledgement}
This research is partially supported by JST ACT-I Grant Number JPMJPR18UD and by JSPS KAKENHI Grant-in-Aid for Young Scientists (B) (JP15K21069), Japan.


\bibliographystyle{ACM-Reference-Format}
\bibliography{paper-65}

\end{document}